\documentclass[a4paper,twocolumn,aps,prb,10pt]{revtex4-2}
\usepackage{graphicx} 
\usepackage{geometry}
\usepackage[english]{babel}
\usepackage{color}
\usepackage[dvipsnames]{xcolor}
\usepackage{textcomp}
\usepackage{amsfonts, amsmath, dsfont}
\usepackage{esint}
\usepackage{xfrac}
\usepackage[unicode, colorlinks=true, allcolors=blue]{hyperref}
\usepackage{dcolumn}
\usepackage{pgfplots}
\pgfplotsset{compat=newest}
\usepgfplotslibrary{groupplots}
\usepackage{tikz}
\usetikzlibrary{shapes.geometric,arrows.meta}
\usetikzlibrary{calc, intersections, plotmarks}
\usetikzlibrary{pgfplots.groupplots}
\usetikzlibrary{backgrounds, positioning, shadows}
\usepackage{caption}
\usepackage{subcaption}
\usepackage{multirow}
\usepackage{arydshln}
\usepackage{booktabs}
\usepackage{tabularx}
\usepackage{siunitx}

\let\oldaddcontentsline\addcontentsline
\renewcommand{\addcontentsline}[3]{}

\begin{document}

\title{Molecular reference corrections for quantum Monte Carlo adsorption energies  }

\author{Roman Fanta}
\email{rfanta@stanford.edu}
\thanks{Corresponding author}
\affiliation{SUNCAT Center for Interface Science and Catalysis, SLAC National Accelerator Laboratory, Menlo Park, California 94025, United States}
\affiliation{Department of Chemical Engineering, Stanford University, Stanford, California 94305, United States}

\author{Michal Bajdich}
\email{bajdich@slac.stanford.edu}
\thanks{Corresponding author}
\affiliation{SUNCAT Center for Interface Science and Catalysis, SLAC National Accelerator Laboratory, Menlo Park, California 94025, United States}

\begin{abstract}
Accurate surface thermochemistry requires balanced error cancellation between extended slabs and molecular reference states. This balance can fail whenever the electronic-structure error is not transferable across the chemically distinct species entering a thermodynamic cycle. Here we examine this problem in single-determinant fixed-node diffusion Monte Carlo (SD-FNDMC) for oxygen-reduction intermediates on Pt(111) and selected CO-reduction intermediates on Cu(111). Gas-phase thermochemistry is used to diagnose the reference-state imbalance, and a hybrid cycle is introduced to separate slab--adsorbate binding from molecular formation. The hybrid cycle keeps the surface binding term at the SD-FNDMC level, where cancellation is expected to be most favorable, and replaces the molecular formation contribution with a benchmark coupled cluster reference. For Pt(111), the resulting correction is small for O and OH but larger for OOH, while the geometry-matched refinement gives only a secondary correction. Applying the same cycle to $^\ast$CHO and $^\ast$COH adsorption on Cu(111) gives corrections of opposite sign, showing that the bias is controlled primarily by the electronic structure of the corresponding HCO and COH molecular references rather than by adsorbate geometry alone. This decomposition identifies molecular reference imbalance as a separable source of error in SD-FNDMC surface thermochemistry and reduces the corresponding bias without modifying the SD-FNDMC slab-binding contribution.
\end{abstract}

\maketitle

\section{Introduction}
Adsorption energies are the electronic quantities that enter most surface reaction free-energy diagrams, and errors of only a few tenths of an eV can change predicted site preferences, limiting potentials, and volcano-plot trends, which relate catalytic activity to the binding strength of key reaction intermediates. On metallic catalysts, this accuracy remains difficult to obtain with standard semilocal density functional theory (DFT), which can give substantial functional dependence for key intermediates \cite{Wellendorff2012,wellendorff2015,araujo2022,kothakonda2026}, while hybrids and other higher-rung approximations are often problematic or impractical for metals and extended slabs \cite{paier2007,stroppa2008,garza2018,urrego-ortiz2024}. The difficulty is amplified in electrocatalytic thermochemistry because the energy of a bound intermediate is usually referenced to gas-phase molecules such as H$_2$O, H$_2$, O$_2$, CO, HCO, or COH. The final adsorption free energy depends not only on the adsorbate--surface interaction but also on how errors cancel between chemically dissimilar molecular and surface-adsorbed states \cite{sargeant2021,urrego-ortiz2024}.

In DFT-based catalysis, this reference-state problem is usually handled by correcting the molecular side of the thermodynamic cycle. Here, a thermodynamic cycle denotes an algebraically equivalent decomposition of the adsorption energy that connects the clean surface and gas-phase molecular references to the adsorbed intermediate. For the oxygen reduction reaction (ORR) and oxygen evolution reaction (OER), these references are commonly constructed from H$_2$ and H$_2$O within the computational hydrogen electrode (CHE) framework \cite{norskov2004, seh2017}. The standard O$_2$ correction used in ORR/OER modeling is the simplest example, and more recent molecule-, bond-, or functional-group-based corrections extend the same idea to broader sets of gas-phase species \cite{granda-marulanda2020, sargeant2021, urrego-ortiz2021, urrego-ortiz2024}. Related DFT studies have identified systematic bond-specific errors by comparing gas-phase and adsorbed reaction energies, including errors associated with C--O double bonds in carbon-containing species and the O--O bond in $^\ast$OOH \cite{Christensen2015,Christensen2016}. More recent work has further separated total adsorption-energy errors into gas-phase and adsorbed-phase contributions for $^\ast$OH, $^\ast$CO, and $^\ast$COOH using combined experimental and computational analyses \cite{Romeo2025,Urregoortiz2025}.

Such corrections can improve adsorption energies, equilibrium potentials, scaling relations, and volcano plots when the dominant error is carried by the gas-phase references \cite{sargeant2021,urrego-ortiz2024,basera2025}. Related reference-energy fitting strategies, such as fitted elemental-phase reference energies, apply the same principle to formation energies by adjusting elemental reference states to improve agreement with experiment \cite{stevanovic2012}. These corrections are nevertheless tied to a particular functional and/or workflow: they adjust the molecular reference energies entering a chosen thermodynamic cycle, but they do not by themselves correct errors in the adsorbate--surface binding contribution when that term is the limiting source of error \cite{urrego-ortiz2024,Romeo2025,Urregoortiz2025}.

These limitations motivate a complementary many-body approach that can be applied directly to realistic extended surfaces without relying on an approximate exchange-correlation functional. Fixed-node diffusion Monte Carlo \cite{ceperley1977,foulkes2001} (FNDMC), a real-space quantum Monte Carlo (QMC) method, provides a many-body route to surface energetics whose accuracy is controlled primarily by the quality of the trial nodal surface \cite{umrigar2007,morales2012,powell2016}. Several studies have already demonstrated high-quality energetics for heterogeneous catalysis and related surface-reaction benchmarks \cite{doblhoff-dier2017,shin2019,iyer2022,stachova2023,shi2023,fanta_CO_2025}.

Although FNDMC provides a promising approach for benchmarking adsorption energetics on extended surfaces, its practical accuracy is limited by the fixed-node approximation, whose residual error is not generally transferable across chemically dissimilar states \cite{morales2012,dubecky2017}. Here, ``not transferable'' does not refer to the absolute total-energy accuracy of an individual state. As in most QMC applications, the relevant issue is how well the residual fixed-node errors cancel in the energy difference of interest. Such cancellation is generally most favorable between electronically similar states and can deteriorate when a thermodynamic cycle connects chemically dissimilar molecular and extended-system states. In the widely used single-determinant Slater-Jastrow (SD) form, this is already evident in molecular benchmarks, where SD-FNDMC shows uneven performance for atomization energies and related total-energy differences \cite{nemec2010,petruzielo2012,wang_performance_2019}. 

Recent studies of noncovalent complexes and $\pi\pi$-stacked dimers show that SD-FNDMC can exhibit systematic biases even for relatively simple energy differences, such as benzene-dimer binding, where no strong bonds are formed or broken and single-reference coupled-cluster methods remain accurate \cite{dubecky2017,Sulka2023,fanta_why_2025}. This demonstrates that small state-dependent nodal errors can become significant when they do not cancel between the compared states, without implying severe multireference character. The relevance to adsorption thermochemistry is therefore not that the ORR/OER intermediates are assumed to be strongly multireference, but that state-dependent fixed-node penalties may fail to cancel when slab-bound intermediates are referenced to chemically dissimilar molecular states with different bonding motifs. The H$_2$O/H$_2$ CHE reference is not intrinsically problematic, but the residual SD-FNDMC errors associated with this reference combination do not necessarily cancel to the same degree against those of different intermediates, as demonstrated below for OH and OOH. Multideterminant trial functions can reduce this residual nodal bias, but at substantially greater cost for realistic surface calculations \cite{morales2012,giner2015,spanedda2025}.

The central aim is to determine whether a distinct molecular reference-state contribution can be identified and treated separately from the residual errors in the SD-FNDMC slab-binding terms. To investigate this, the present work first examines gas-phase oxygen and hydroxyl species relevant to the ORR, and subsequently develops adsorption-energy schemes that separate the contributions of slab binding, molecular formation, and adsorbate relaxation.

The resulting cycle is ``hybrid'' because it combines two electronic-structure methods within the same thermodynamic decomposition: the extended-system binding contribution is evaluated with SD-FNDMC, whereas the molecular formation term and, when needed, the relaxation term are evaluated with a high-level molecular benchmark method. Although demonstrated here for adsorption energies on Pt(111) and Cu(111), the same decomposition applies when the target thermochemical quantity contains an extended-system contribution and a separately benchmarkable molecular reference term. Its purpose is to diagnose and reduce the additional bias introduced when SD-FNDMC compares extended-system states to chemically dissimilar molecular references, while leaving any residual fixed-node error in the SD-FNDMC contribution unchanged.

Here, oxygenated intermediates on Pt(111) serve as the primary test case because Pt(111) is the canonical benchmark surface for ORR \cite{norskov2004,rossmeisl2005,seh2017}. We first use gas-phase O/OH/OOH thermochemistry to diagnose the reference-state imbalance in SD-FNDMC, and then compare three electronic-energy constructions for the Pt(111) intermediates: the conventional CHE scheme for ORR, a reference-state-balanced hybrid cycle with optimized isolated adsorbates, and a geometry-matched refinement for OOH. Transferability is then tested using previously published SD-FNDMC adsorption data for $^\ast$CHO and $^\ast$COH on Cu(111) \cite{fanta_CO_2025}. Here and throughout, $^\ast$ denotes a surface site and $^\ast X$ denotes species $X$ adsorbed on the surface. The corresponding isolated radicals are denoted HCO (formyl) and COH (isoformyl), respectively. Representative adsorbed geometries are reported in Ref.~\cite{fanta_CO_2025}. The present work adds only the molecular reference calculations required for the hybrid cycle analysis. This sequence separates the diagnostic gas-phase benchmark, the adsorption-energy correction, and the cross-chemistry transferability test.

\section{Methods}
\subsection{DFT structural models and geometry optimization}
Structural models for the bulk, slabs, adsorbates, and isolated molecular references were relaxed in the Vienna \textit{ab initio} Simulation Package (VASP) \cite{kresse1993ab,kresse1994ab,kresse1996efficient,kresse1996efficiency,kresse1999ultrasoft} at the spin-polarized PBE-D3(BJ) \cite{Perdew1996,D3,D3BJ} level. The recommended PBE projector-augmented wave (PAW) potentials were used together with a plane-wave cutoff of 520~eV, Gaussian smearing of 0.05~eV, and $\Gamma$-centered $k$-point meshes. For slab and molecular optimizations, the electronic and ionic convergence criteria were set to $10^{-6}$~eV and $0.02$~eV/\AA, respectively. Bulk Pt was optimized more tightly, using convergence thresholds of $10^{-8}$~eV and $0.002$~eV/\AA. Ionic relaxation was performed with the conjugate-gradient algorithm.

The fcc Pt lattice constant optimized at this level was 3.93~\AA\ using a $16\times16\times16$ $k$-point mesh. PBE-D3(BJ) was retained as a common geometry-generation protocol for all subsequent electronic-energy calculations. Using separately optimized structures for each method would introduce method-dependent geometric differences as an additional variable, whereas the present study is designed to compare the electronic-energy contributions on a single common structure.

The Pt(111) slab was then constructed as a three-layer slab containing 12 Pt atoms in total (four per layer, corresponding to a Pt(111)-(\(2\times2\)) surface cell and \(1/4\)~ML coverage for a single adsorbate), with cell dimensions of 5.55, 5.55, and 20.54~\AA. For the finite-size analysis, this 12-atom slab is denoted the \(1\times1\) QMC simulation cell. Its \(2\times2\) tiling contains 48 Pt atoms and four adsorbates, thereby preserving the \(1/4\)-ML coverage. The bottom Pt layer was fixed at the bulk positions, while the upper two layers and all adsorbates were allowed to relax. Slab calculations employed a $12\times12\times1$ $\Gamma$-centered $k$-point mesh and approximately 16~\AA\ of vacuum between periodic images. Isolated molecular reference species were placed in a $12\times12\times12$~\AA$^3$ supercell and sampled at the $\Gamma$ point.

\subsection{Gas-phase thermochemistry}
For the gas-phase DFT thermochemistry benchmarks, single-point calculations were carried out in ORCA \cite{ORCA,ORCA_4.0,ORCA_5.0} on the VASP-relaxed geometries. The ORCA calculations employed the def2-TZVPPD \cite{weigend2005a,rappoport2010a} basis set together with the D4 dispersion correction \cite{D4,D4_extension}. Closed-shell species (H$_2$O, H$_2$, and CO) were treated with restricted Kohn--Sham (RKS), whereas open-shell species (OH, OOH, O$_2$, HCO, and COH) were treated with unrestricted Kohn--Sham (UKS) using the appropriate spin multiplicity in each case. Wave-function stability of the open-shell solutions was checked by stability analysis. The benchmarked exchange-correlation functionals were PBE \cite{perdew1996pbe}, RPBE \cite{Hammer1999}, SCAN \cite{Sun2015scan}, PBE0 \cite{adamo1999}, and B3LYP \cite{stephens1994b3lyp}. In addition, BEEF-vdW \cite{Wellendorff2012}, which is not available in ORCA, was evaluated in VASP using the same computational settings as the PBE-D3(BJ) calculations. ORCA was chosen for these molecular benchmarks because it provides direct control over restricted and unrestricted open-shell calculations and built-in stability analysis of the Kohn--Sham solutions. Each thermochemical reaction energy was constructed using total energies obtained with a single functional and code, so ORCA and VASP total energies were not combined within any reaction. The BEEF-vdW results are included only in the comparative gas-phase benchmark and do not enter Schemes~1 or 2.

Calculated thermochemical quantities at 298.15~K were obtained by adding zero-point energies and integrated heat capacities from 0 to 298.15~K to the electronic energies. These corrections were taken from the experimental part of NIST Computational Chemistry Comparison and Benchmark Database (CCCBDB) \cite{CCCBDB2022}, and the corresponding 298.15~K formation enthalpies were constructed consistently for all molecular reference species. Experimental reference data were taken from standard thermochemical compilations \cite{Gurvich1989,Cox1989}, and targeted literature sources \cite{Ramond2002,Ruscic2006} for specific radicals such as OOH. The detailed molecular thermochemical data, including zero-point energies, integrated heat capacities, and the resulting 298.15~K formation enthalpies, are provided in the SI.

\subsection{Coupled cluster molecular reference calculations}
Reference molecular energies were evaluated in MRCC \cite{MRCC, mrcc_website} using the same underlying geometries. Restricted or unrestricted Hartree--Fock references were used consistently with the corresponding molecular spin state, and all coupled cluster (CC) calculations employed the frozen-core approximation. CCSD(T) \cite{Raghavachari1989,rolik2013,gyevi-nagy2020} and CCSDT(Q) \cite{bomble2005,kallay2005,kallay2008} total energies were extrapolated to the complete-basis set limit using a two-point extrapolation \cite{Helgaker1997,halkier1998} from cc-pVDZ and cc-pVTZ \cite{Dunning1989} calculations 
\begin{equation}
E_{\mathrm{CBS}}
= \frac{3^3 E_{\mathrm{VTZ}} - 2^3 E_{\mathrm{VDZ}}}{3^3-2^3}.
\end{equation}
Small differences between CCSD(T)/CBS and CCSDT(Q)/CBS results were used to assess the adequacy of CCSD(T)/CBS as the molecular benchmark values entering the hybrid correction schemes.

\subsection{QMC orbital generation and trial wave functions}
Single-particle orbitals used in the QMC calculations were obtained from Quantum Espresso \cite{Giannozzi2009,Giannozzi2017,Giannozzi2020} PBE calculations using norm-conserving correlation-consistent effective core potentials (ccECPs) designed for QMC simulations \cite{Bennett2017, Annaberdiyev2018, Zhou2024}. Spin treatment was chosen according to the molecular or adsorbate spin state, and all calculations used the VASP-relaxed geometries without further structural relaxation. A kinetic-energy cutoff of 400~Ry and Gaussian smearing of $3.68\cdot10^{-3}$~Ry were used for the orbital-generation step. For the Pt(111) slab calculations, the orbitals were generated with a $12\times12\times1$ $k$-point mesh for the base \(1\times1\) QMC simulation cell, while isolated molecular references were treated at the $\Gamma$ point in a large $12\times12\times12$~\AA$^3$ supercell.

All many-body calculations were carried out in QMCPACK \cite{Kim2018,Kent2020} within a single-determinant Slater--Jastrow approach. The Slater determinant employed the PBE orbitals from Quantum Espresso, which were converted from the plane-wave representation to a three-dimensional B-spline basis in QMCPACK. The Jastrow factor always included one-body, two-body, and three-body terms (J1+J2+J3), corresponding to electron--nucleus, electron--electron, and electron--electron--nucleus correlations \cite{drummond2004}. The Jastrow parameters were optimized within variational Monte Carlo using the linear method \cite{Toulouse2007}.

\subsection{Diffusion Monte Carlo sampling and finite-size extrapolation}
The final energies were obtained from fixed-node diffusion Monte Carlo runs with a time step of $\Delta\tau=0.005$ a.u., using the variational T-moves \cite{Casula2006,Casula2010} treatment of the nonlocal pseudopotentials. No time-step extrapolation was performed. The choice $\Delta\tau=0.005$~a.u. was guided by earlier benchmarks \cite{doblhoff-dier2016} and by the high acceptance ratios observed in our production calculations. Many-body finite-size effects were treated solely through twist averaging \cite{lin2001} and extrapolation with cell size \cite{Drummond2008,iyer2022}. 
Twist averaging was carried out with a $12\times12\times1$ twist grid for the base \(1\times1\) QMC cell and a $6\times6\times1$ twist grid for its \(2\times2\) tiling, corresponding to the same twist density. Typical walker populations were approximately $10^4$ per twist for the $1\times1$ slabs and for isolated molecules, and approximately $2\times10^4$ per twist for the $2\times2$ slabs. For isolated molecules, this corresponds to the $\Gamma$ point only.

Residual finite-size effects in adsorption energies were removed by $N^{-5/4}$ extrapolation \cite{Drummond2008} between the base \(1\times1\) QMC cell and its \(2\times2\) tiling
\begin{equation}
\Delta E_{\infty}
= \frac{N_2^{-5/4}\Delta E_1 - N_1^{-5/4}\Delta E_2}{N_2^{-5/4}-N_1^{-5/4}},
\end{equation}
where \(N_1\) and \(N_2\) denote the numbers of electrons in the two simulation cells and \(\Delta E_1\) and \(\Delta E_2\) are the corresponding adsorption energies. All electronic-structure and QMC inputs were assembled through the Nexus workflow infrastructure \cite{Krogel2016}.

For the Cu(111) transferability analysis, the previously published slab geometries and SD-FNDMC adsorption data for $^\ast$CO, $^\ast$COH, and $^\ast$CHO from Ref.~\cite{fanta_CO_2025} were reused. In the present work, only the additional molecular reference calculations needed for the hybrid-cycle analysis were carried out, namely the gas-phase HCO and COH benchmarks and the isolated frozen HCO/COH radicals extracted from the adsorbed Cu(111) geometries. The underlying Cu(111) surface structures and electronic adsorption data are available from Ref.~\cite{fanta_CO_2025}.

\section{Results and discussion}
\subsection{Gas-phase thermochemistry reveals the reference-state problem}
The first step in the molecular correction scheme is to determine whether SD-FNDMC treats the oxygen-derived molecular references entering ORR energetics with comparable accuracy. This is essential because any adsorption cycle ultimately inherits its accuracy from the reference formulations used to define it. Because ORR energetics are commonly analyzed within the computational hydrogen electrode (CHE) method introduced by N{\o}rskov and co-workers \cite{norskov2004}, we explicitly consider CHE-derived reference formulations. In the CHE model, the free energy of a transferred proton--electron pair is referenced to gas-phase H$_2$, allowing electrochemical free energies to be constructed from neutral species and shifted with applied potential. To test how this reference choice affects SD-FNDMC error cancellation, we compare direct molecular formation enthalpies for H$_2$O, OH, and OOH with the corresponding CHE-derived reference quantities for OH and OOH. All five thermochemical quantities were computed using zero-point and integrated heat capacity corrections and compared against experimental values and high-level coupled cluster benchmarks. If SD-FNDMC errors were transferable across these reference formulations, all five quantities would show comparable deviations from experiment. Instead, the results show a strongly reference-dependent pattern, indicating that the molecular reference construction itself is part of the problem addressed here.

It is convenient to define
\begin{equation}
\widetilde{E}_X = E_X +  E_X^{\mathrm{ZPE}} + \Delta H_X^{0\rightarrow 298.15~\mathrm{K}},
\label{eq:enthalpy}
\end{equation}
where
$$
\Delta H_X^{0\rightarrow298.15~\mathrm{K}}=H_X^\circ(298.15~\mathrm{K})-H_X^\circ(0~\mathrm{K}),
$$
such that the direct molecular formation enthalpies are
\begin{align}
dH_{\mathrm{H_2O}} &= \widetilde{E}_{\mathrm{H_2O}} - \widetilde{E}_{\mathrm{H_2}} - \frac{1}{2}\widetilde{E}_{\mathrm{O_2}}, \label{eq:h2o_enthalpy}\\
dH_{\mathrm{OH}}   &= \widetilde{E}_{\mathrm{OH}} - \frac{1}{2}\widetilde{E}_{\mathrm{H_2}} - \frac{1}{2}\widetilde{E}_{\mathrm{O_2}}, \label{eq:oh_enthalpy}\\
dH_{\mathrm{OOH}}  &= \widetilde{E}_{\mathrm{OOH}} - \frac{1}{2}\widetilde{E}_{\mathrm{H_2}} - \widetilde{E}_{\mathrm{O_2}},\label{eq:ooh_enthalpy}
\end{align}
whereas the CHE-derived formulations are
\begin{align}
dH_{\mathrm{OH}}^{\mathrm{CHE}}  &= \widetilde{E}_{\mathrm{OH}} - \widetilde{E}_{\mathrm{H_2O}} + \frac{1}{2}\widetilde{E}_{\mathrm{H_2}}, \label{eq:che_oH_enthalpy}\\
dH_{\mathrm{OOH}}^{\mathrm{CHE}} &= \widetilde{E}_{\mathrm{OOH}} - 2\widetilde{E}_{\mathrm{H_2O}} + \frac{3}{2}\widetilde{E}_{\mathrm{H_2}}. \label{eq:che_ooh_enthalpy}
\end{align}

Atomic O is not included as a separate entry in the gas-phase thermochemistry diagnostic because its direct O$_2$-based formation is equivalent to one half of the O$_2$ dissociation or atomization enthalpy. This quantity mainly probes the O$_2$/O atom splitting rather than the molecular-reference cancellation channels emphasized here. The atomic O reference is nevertheless included in the adsorption-energy schemes through the CHE-based O$^\ast$ reference, \( \mu_{\mathrm{O}}^{\mathrm{CHE}} = E_{\mathrm{H_2O}} - E_{\mathrm{H_2}} \), which gives the Scheme~1 correction reported below.

\begin{figure*}[htbp]
\centering
\definecolor{cPBE}{HTML}{81007f}
\definecolor{cRPBE}{HTML}{9A4058}
\definecolor{cBEEF}{HTML}{f55f74}
\definecolor{cSCAN}{HTML}{a672eb}
\definecolor{cPBE0}{HTML}{226A4D}
\definecolor{cB3LYP}{HTML}{77BE20}
\definecolor{cCCSD}{HTML}{FFC326}
\definecolor{cCCSDT}{HTML}{fd8d3c}
\definecolor{cCCSDTQ}{HTML}{e31a1c}
\definecolor{cFNDMC}{HTML}{000000}

\begin{tikzpicture}
\begin{axis}[
    width=\textwidth,
    height=7.5cm,
    enlarge x limits=0.1,
    symbolic x coords={H$_2$O, OH, OOH, OH(CHE), OOH(CHE)},
    xtick=data,
    ytick distance=0.1,
    ylabel={Signed error (eV)},
    xlabel={Species / Reaction},
    ymajorgrids=true,
    ymax=0.45,
    ymin=-0.75,
    grid style={gray!30, dashed},
    tick label style={font=\footnotesize},
    label style={font=\small},
    extra y ticks={0},
    extra y tick style={grid=major, grid style={thick, black}},
    legend columns=5,
    legend style={
        draw=gray!50,
        fill=white,
        font=\footnotesize,
        at={(0.5,1.05)},
        anchor=south,
        /tikz/every even column/.append style={column sep=0.5cm}
    }
]
\addplot+[
    color=cPBE!85, mark=square*, thick, solid, mark options={solid, fill=cPBE!85}
] coordinates {
    (H$_2$O,0.201) (OH,0.275) (OOH,0.004) (OH(CHE),0.074) (OOH(CHE),-0.397)
};
\addlegendentry{PBE}

\addplot+[
    color=cRPBE, mark=triangle*, thick, solid, mark options={solid, fill=cRPBE}
] coordinates {
    (H$_2$O,0.342) (OH,0.233) (OOH,0.087) (OH(CHE),-0.109) (OOH(CHE),-0.597)
};
\addlegendentry{RPBE}

\addplot+[
    color=cBEEF, mark=diamond*, thick, solid, mark options={solid, fill=cBEEF}
] coordinates {
    (H$_2$O,0.387) (OH,0.164) (OOH,0.082) (OH(CHE),-0.223) (OOH(CHE),-0.692)
};
\addlegendentry{BEEF-vdW}

\addplot+[
    color=cSCAN, mark=pentagon*, thick, solid, mark options={solid, fill=cSCAN}
] coordinates {
    (H$_2$O,0.172) (OH,0.044) (OOH,0.016) (OH(CHE),-0.128) (OOH(CHE),-0.328)
};
\addlegendentry{SCAN}

\addplot+[
    color=cPBE0, mark=*, thick, solid, mark options={solid, fill=cPBE0}
] coordinates {
    (H$_2$O,0.085) (OH,0.042) (OOH,0.025) (OH(CHE),-0.044) (OOH(CHE),-0.146)
};
\addlegendentry{PBE0}

\addplot+[
    color=cB3LYP, mark=square*, thick, solid, mark options={solid, fill=cB3LYP}
] coordinates {
    (H$_2$O,0.167) (OH,0.045) (OOH,0.035) (OH(CHE),-0.122) (OOH(CHE),-0.299)
};
\addlegendentry{B3LYP}
\addplot+[
    color=cCCSD, mark=triangle*, thick, dashed, mark options={solid, fill=white}
] coordinates {
    (H$_2$O,-0.025) (OH,-0.113) (OOH,-0.018) (OH(CHE),-0.087) (OOH(CHE),0.032)
};
\addlegendentry{CCSD}

\addplot+[
    color=cCCSDT, mark=diamond*, thick, dashed, mark options={solid, fill=white}
] coordinates {
    (H$_2$O,0.001) (OH,-0.004) (OOH,-0.014) (OH(CHE),-0.005) (OOH(CHE),-0.015)
};
\addlegendentry{CCSD(T)}

\addplot+[
    color=cCCSDTQ, mark=pentagon*, thick, dashed, mark options={solid, fill=white}
] coordinates {
    (H$_2$O,0.014) (OH,0.005) (OOH,-0.018) (OH(CHE),-0.009) (OOH(CHE),-0.046)
};
\addlegendentry{CCSDT(Q)}
\addplot+[
    color=cFNDMC, mark=*, thick, dotted, mark options={solid, fill=cFNDMC}
] coordinates {
    (H$_2$O,-0.097) (OH,-0.065) (OOH,-0.026) (OH(CHE),0.032) (OOH(CHE),0.168)
};
\addlegendentry{SD-FNDMC}

\draw[very thick, dashed, gray] (rel axis cs:0.604, 0) -- (rel axis cs:0.604, 1);

\end{axis}
\end{tikzpicture}

\caption{Signed errors relative to the experimental reference values} for 298.15~K gas-phase thermochemistry of H$_2$O, OH, and OOH, together with the corresponding CHE-derived OH and OOH quantities, in eV. Negative values indicate that the calculated thermochemical quantity is lower than the experimental value, while positive values indicate that it is higher. The corresponding calculated values are reported in Table~\ref{tab:thermochem_errors}.
\label{fig:thermochem_errors}
\end{figure*}
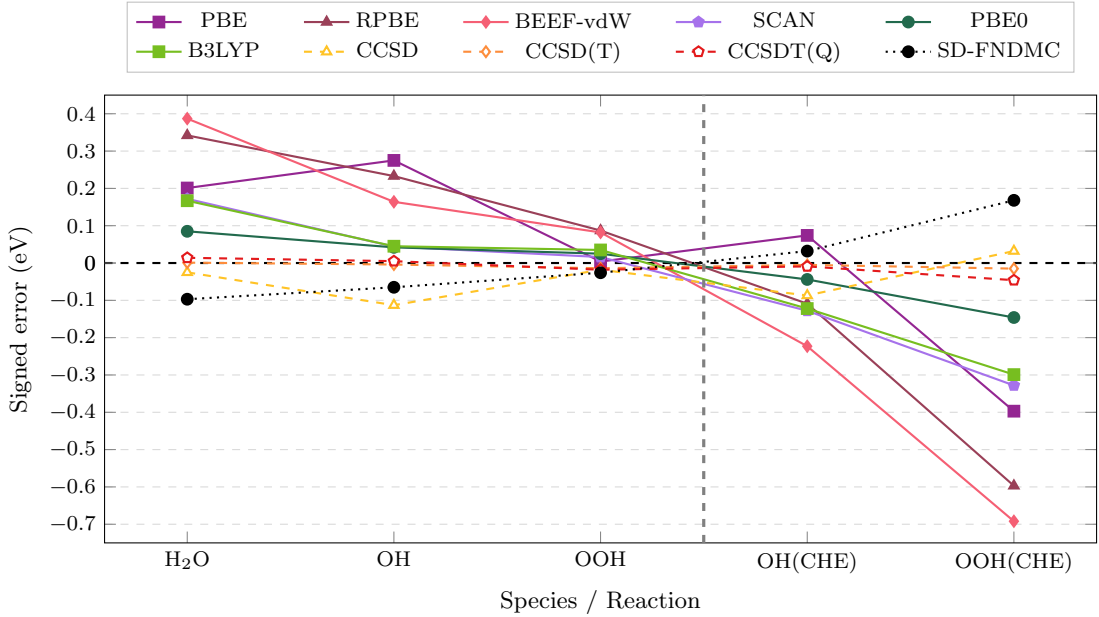

The signed-error summary in Fig.~\ref{fig:thermochem_errors} makes the reference-state dependence immediately visible: methods that are reasonably accurate for direct OOH formation can still show substantially larger deviations for the CHE-derived OOH quantity, indicating that the thermodynamic cycle itself contributes to the final error. The corresponding calculated values at 298.15~K and signed deviations from experiment are reported in Table~\ref{tab:thermochem_errors}. For each method, electronic energies were combined with the same zero-point and integrated heat capacity corrections to construct the corresponding 298.15~K enthalpy-like quantities. The signed error is defined as
\begin{equation}
\varepsilon_i = dH_i^{\mathrm{calc}} - dH_i^{\mathrm{exp}},
\label{eq:enthalpy_err}
\end{equation}
so that negative values indicate that the calculated thermochemical quantity is lower than experiment (more negative, or less positive), whereas positive values indicate that it is higher (less negative, or more positive). In Table~\ref{tab:thermochem_errors}, each entry is given as the calculated reaction energy, together with the corresponding signed error in brackets. The tabulated quantities correspond directly to the reactions as written and are not normalized per O atom or per O$_2$ molecule. Accordingly, the H$_2$O and OH formation entries correspond to reactions containing $\tfrac{1}{2}$O$_2$, whereas the OOH formation entry corresponds to a reaction containing one O$_2$ molecule.

Density-functional results already show that this reference-state dependence is not unique to QMC. Across the semilocal functionals, the H$_2$O and OH formation enthalpies are typically higher than experiment, whereas the direct OOH formation error is often much smaller. This does not contradict the original motivation of CHE in DFT, where replacing explicit O$_2$-based electrochemical steps by H$_2$/H$_2$O-based references often improves robustness. It does, however, show that the most suitable reference construction remains method- and intermediate-dependent. For OOH, the CHE-derived quantity probes a different and less favorable cancellation channel: in direct formation, the OOH error is balanced against O$_2$, whereas in the CHE-derived expression, it is balanced against the H$_2$O/H$_2$ reference. Thus, even if H$_2$O and H$_2$ are individually more accurate than O$_2$, they do not necessarily provide a better error match to OOH. The hybrid functionals show the same qualitative behavior, although with smaller absolute deviations.

SD-FNDMC exhibits the same general sensitivity to reference construction, but in a more revealing form. Its errors are moderate for H$_2$O and OH formation, small for direct OOH formation, but substantially larger for CHE cycle OOH. In particular, the OOH formation enthalpy referenced directly to $\tfrac{1}{2}$H$_2$~+~O$_2$ is close to experiment, whereas the CHE-derived OOH quantity shows a much larger deviation. The direct formation errors further suggest that the O$_2$ reference plays a special role in the SD-FNDMC thermochemistry. Because H$_2$ is almost exact in SD-FNDMC, the corresponding signed errors can be written approximately as
\begin{align*}
\varepsilon_{\mathrm{H_2O}} \approx \delta_{\mathrm{H_2O}} & -\tfrac{1}{2}\delta_{\mathrm{O_2}},
\quad \varepsilon_{\mathrm{OH}} \approx \delta_{\mathrm{OH}}-\tfrac{1}{2}\delta_{\mathrm{O_2}},\\
&\varepsilon_{\mathrm{OOH}} \approx \delta_{\mathrm{OOH}}-\delta_{\mathrm{O_2}},
\end{align*}
where $\delta_X$ denotes the residual SD-FNDMC bias of species $X$. Uniformly negative direct-formation errors are consistent with an O$_2$ reference whose residual bias is larger than that of H$_2$O, OH, and OOH on the corresponding stoichiometric scale. In this picture, the comparatively small direct OOH error reflects favorable cancellation between the residual biases of OOH and O$_2$, whereas the larger negative H$_2$O and OH errors indicate that this cancellation is less complete for those species.

\begin{table*}[htbp] 
\caption{
Calculated reaction energies and signed errors [in brackets] for 298.15~K molecular thermochemistry, in eV. Direct formation reactions for H$_2$O and OH contain $\tfrac{1}{2}$O$_2$, whereas the OOH reaction contains one O$_2$.
}
\label{tab:thermochem_errors}
\begin{tabular*}{\textwidth}{@{\extracolsep{\fill}} l *{5}{S[table-format=-1.3, table-space-text-post={~[-0.000]}]} @{}} 
\toprule\toprule
& \multicolumn{3}{c}{\textbf{Direct formation}} & \multicolumn{2}{c}{\textbf{CHE-derived}} \\
\cmidrule(lr){2-4} \cmidrule(lr){5-6}
\textbf{Method} & 
\multicolumn{1}{c}{\textbf{H$_2$O}} & 
\multicolumn{1}{c}{\textbf{OH}} & 
\multicolumn{1}{c}{\textbf{OOH}} & 
\multicolumn{1}{c}{\textbf{OH(CHE)}} & 
\multicolumn{1}{c}{\textbf{OOH(CHE)}} \\ 
\midrule
PBE 		 & -2.305 {~[ 0.201]} & 0.662 {~[ 0.275]} & 0.132 {~[ 0.004]} & 2.967 {~[ 0.074]} & 4.743 {~[-0.397]} \\
RPBE 		 & -2.164 {~[ 0.342]} & 0.620 {~[ 0.233]} & 0.215 {~[ 0.087]} & 2.784 {~[-0.109]} & 4.543 {~[-0.597]} \\
BEEF-vdW 	 & -2.119 {~[ 0.387]} & 0.551 {~[ 0.164]} & 0.210 {~[ 0.082]} & 2.670 {~[-0.223]} & 4.448 {~[-0.692]} \\
SCAN 		 & -2.334 {~[ 0.172]} & 0.431 {~[ 0.044]} & 0.144 {~[ 0.016]} & 2.765 {~[-0.128]} & 4.812 {~[-0.328]} \\
PBE0 		 & -2.421 {~[ 0.085]} & 0.429 {~[ 0.042]} & 0.153 {~[ 0.025]} & 2.849 {~[-0.044]} & 4.994 {~[-0.146]} \\
B3LYP 		 & -2.339 {~[ 0.167]} & 0.432 {~[ 0.045]} & 0.163 {~[ 0.035]} & 2.771 {~[-0.122]} & 4.841 {~[-0.299]} \\
CCSD 		 & -2.531 {~[-0.025]} & 0.274 {~[-0.113]} & 0.110 {~[-0.018]} & 2.806 {~[-0.087]} & 5.172 {~[ 0.032]} \\
CCSD(T) 	 & -2.505 {~[ 0.001]} & 0.383 {~[-0.004]} & 0.114 {~[-0.014]} & 2.888 {~[-0.005]} & 5.125 {~[-0.015]} \\
CCSDT(Q) 	 & -2.492 {~[ 0.014]} & 0.392 {~[ 0.005]} & 0.110 {~[-0.018]} & 2.884 {~[-0.009]} & 5.094 {~[-0.046]} \\
SD-FNDMC 	 & -2.603 {~[-0.097]} & 0.322 {~[-0.065]} & 0.102 {~[-0.026]} & 2.925 {~[ 0.032]} & 5.308 {~[ 0.168]} \\
\midrule
\textbf{Experiment} & 
\multicolumn{1}{c}{$-2.506$} & 
\multicolumn{1}{c}{$0.387$} & 
\multicolumn{1}{c}{$0.128$} & 
\multicolumn{1}{c}{$2.893$} & 
\multicolumn{1}{c}{$5.140$} \\ 
\bottomrule\bottomrule
\end{tabular*}
\end{table*}

The CHE cycle OOH error probes a different cancellation channel:
\[
\varepsilon_{\mathrm{OOH}}^{\mathrm{dir}} \approx \delta_{\mathrm{OOH}}-\delta_{\mathrm{O_2}},
\quad
\varepsilon_{\mathrm{OOH}}^{\mathrm{CHE}} \approx \delta_{\mathrm{OOH}}-2\delta_{\mathrm{H_2O}}.
\]
A near-zero error for direct OOH formation indicates favorable cancellation between the residual SD-FNDMC biases of OOH and O$_2$ in that cycle, not that either species is individually exact. A larger CHE cycle error shows that this favorable cancellation is lost when O$_2$ is replaced by the H$_2$O/H$_2$ reference. These results identify the molecular reference states entering the thermodynamic cycle as a separate source of bias that can be tested independently. This bias can be reduced through an alternative construction of adsorption energies in which the slab-adsorbate binding term is evaluated with QMC, while the molecular formation and, when needed, relaxation terms are evaluated with a high-level benchmark method such as CC.

\subsection{CHE scheme: Conventional adsorption thermochemistry}
Following the gas-phase analysis above, this study first considers a conventional adsorption-energy construction in which SD-FNDMC is used not only for the slab energies, but also for the molecular reference entering the thermodynamic cycle. For an adsorbed intermediate $X$ on a clean surface $^\ast$, the electronic adsorption energy is written as
\begin{equation}
\Delta E_{\mathrm{ads}}^{(0)}(X)
= E_{^\ast X}^{\mathrm{QMC}} - E_{\ast}^{\mathrm{QMC}} - \mu_X^{\mathrm{ref,QMC}},
\label{eq:scheme0_general}
\end{equation}
where $\mu_X^{\mathrm{ref,QMC}}$ is the reference chemical potential of the adsorbate constructed from isolated gas-phase species, evaluated here at the same SD-FNDMC level.

For the oxygenated ORR intermediates considered on Pt(111), the most common choice is the CHE-based reference built from H$_2$O and H$_2$ \cite{norskov2004}. For bookkeeping, this baseline CHE construction is denoted by the superscript (0) in the equations. The corresponding explicit expressions for OH, O, and OOH, together with the analogous direct O$_2$-based reference forms, are given in the Supplemental Material. The essential feature of the CHE scheme remains unchanged: the adsorption energy retains the full reference-state dependence of gas-phase molecular energetics. 

Equation~\eqref{eq:scheme0_general} defines the natural baseline for the adsorption energy analysis because it constructs the full thermodynamic cycle from SD-FNDMC total energies alone. This direct SD-FNDMC cycle is also the most sensitive to reference-state imbalance, because it requires the same fixed-node approximation to describe both the slab-bound intermediate and the chemically distinct gas-phase molecular references. The gas-phase benchmark above already showed that these molecular reference formulations do not exhibit uniform error cancellation. The CHE scheme therefore provides a direct test of whether gas-phase reference errors carry over to surface adsorption.

To compare these computational schemes, only the electronic adsorption energies are considered. Zero-point, thermal, solvation, and electrochemical corrections are omitted from the present comparison. These contributions are not assumed to be independent of the electronic-structure method and would need to be evaluated consistently in a complete free-energy treatment. Accordingly, the differences discussed below isolate only the electronic-energy consequences of the three adsorption-energy constructions.

\begin{table*}[htbp]
\caption{Finite-size-extrapolated adsorption energies on Pt(111) from the CHE scheme, the reference-state-balanced hybrid cycle (Scheme~1), and the geometry-matched refinement (Scheme~2), in eV. Values in parentheses denote the statistical uncertainty in the last digit(s). Scheme~2 was evaluated only for OOH.
}
\label{tab:schemes_pt111}
\begin{tabular*}{\textwidth}{@{\extracolsep{\fill}} l *{3}{S[table-format=1.3(2)]} @{}}
\toprule\toprule
\textbf{Adsorbate/site} &
\multicolumn{1}{c}{\textbf{CHE scheme}} &
\multicolumn{1}{c}{\textbf{Scheme~1}} &
\multicolumn{1}{c}{\textbf{Scheme~2}} \\
\midrule
O$_{\mathrm{fcc}}$   & 1.258(26) & 1.168(30) & \multicolumn{1}{c}{---} \\
O$_{\mathrm{hcp}}$   & 2.188(23) & 2.098(26) & \multicolumn{1}{c}{---} \\
O$_{\mathrm{top}}$   & 3.737(23) & 3.647(29) & \multicolumn{1}{c}{---} \\
OH$_{\mathrm{fcc}}$  & 0.992(22) & 0.955(33) & \multicolumn{1}{c}{---} \\
OH$_{\mathrm{top}}$  & 1.042(32) & 1.006(35) & \multicolumn{1}{c}{---} \\
OOH$_{\mathrm{top}}$ & 4.557(41) & 4.374(44) & 4.333(44) \\
\bottomrule\bottomrule
\end{tabular*}
\end{table*}

\subsection{Scheme 1: Hybrid cycle with optimized isolated adsorbate}
To reduce the reference-state imbalance identified above, a hybrid construction is introduced, in which QMC is retained only for the slab--adsorbate binding contribution, while the molecular formation term is evaluated using a separate high-level benchmark method. The starting point is the isolated adsorbate $X$ in its optimized gas-phase geometry, denoted $X^{\mathrm{opt}}$. This state is distinct from $X^{\mathrm{frz}}$, introduced below in Scheme 2, which denotes the isolated adsorbate frozen in the geometry extracted from the adsorbed structure. Scheme~1 is then defined as
\begin{equation}
\Delta E_{\mathrm{ads}}^{(1)}(X)
\equiv \Delta E_{\mathrm{bind}}^{\mathrm{QMC}}(X^{\mathrm{opt}})
+ \Delta E_{\mathrm{form}}^{\mathrm{HL}}(X^{\mathrm{opt}}),
\label{eq:scheme1_general}
\end{equation}
with
\begin{equation}
\Delta E_{\mathrm{bind}}^{\mathrm{QMC}}(X^{\mathrm{opt}})
= E_{^\ast X}^{\mathrm{QMC}} - E_{\ast}^{\mathrm{QMC}}
- E_{X^{\mathrm{opt}}}^{\mathrm{QMC}},
\label{eq:scheme1_bind}
\end{equation}
and
\begin{equation}
\Delta E_{\mathrm{form}}^{\mathrm{HL}}(X^{\mathrm{opt}})
= E_{X^{\mathrm{opt}}}^{\mathrm{HL}}
- \mu_X^{\mathrm{ref,HL}}.
\label{eq:scheme1_form}
\end{equation}
Here ``HL'' denotes the chosen high-level molecular benchmark, such as CCSD(T)/CBS or a comparably accurate reference method.

Scheme~1 has a simple interpretation. The first term isolates the energy required to bind an already-formed adsorbate to the surface, where QMC error cancellation is expected to be more favorable. The second term replaces the problematic molecular formation energy with a benchmark-quality reference that is not limited by the SD-FNDMC nodal structure of the isolated gas-phase species.

To quantify the numerical effect of Scheme~1, the hybrid construction is applied to the finite-size extrapolated adsorption energies on Pt(111). Because Scheme~1 replaces only the molecular formation term, while leaving the QMC slab--adsorbate binding contribution unchanged, the difference between Schemes~1 and CHE is a site-independent shift for a given adsorbate,
\begin{align}
\begin{split}
\delta_{\mathrm{hyb}}(X)
&=
\Delta E_{\mathrm{ads}}^{(1)}(X)
-
\Delta E_{\mathrm{ads}}^{(0)}(X)\\
&=
\Delta E_{\mathrm{form}}^{\mathrm{HL}}(X)
-
\Delta E_{\mathrm{form}}^{\mathrm{QMC}}(X).
\label{eq:delta_hyb}
\end{split}
\end{align}

Using the CHE-based molecular formation energies obtained above results in
\begin{align*}
\delta_{\mathrm{hyb}}(\mathrm{O}) &= -0.091 \pm 0.010~\mathrm{eV}, \\
\delta_{\mathrm{hyb}}(\mathrm{OH}) &= -0.037 \pm 0.009~\mathrm{eV}, \\
\delta_{\mathrm{hyb}}(\mathrm{OOH}) &= -0.183 \pm 0.014~\mathrm{eV}.
\end{align*}
The uncertainties correspond to one standard deviation and were propagated from the stochastic SD-FNDMC molecular energies; no statistical uncertainty was assigned to the coupled cluster reference energies.

Thus, Scheme~1 lowers the conventional SD-FNDMC adsorption energies modestly for O and OH, but substantially more for OOH. This mirrors the gas-phase thermochemistry benchmark above, where the largest reference-state imbalance was likewise found for the OOH CHE cycle. The correction is therefore not arbitrary but is directly inherited from the molecular reference mismatch, which is diagnosed independently of the surface.

Because the isolated adsorbate enters as the relaxed gas-phase structure $X^{\mathrm{opt}}$, any relaxation required to transform the free molecule into its adsorbed geometry is still implicitly folded into the QMC binding term. Scheme~1 reduces the reference-state problem identified above, but does not yet isolate the geometry mismatch between the free and adsorbed adsorbate. That residual issue motivates the geometry-matched construction introduced in Scheme~2.

\subsection{Scheme 2: Geometry-matched hybrid cycle}
Scheme~2 refines the hybrid construction by introducing the isolated adsorbate frozen in its adsorbed geometry, denoted \(X^{\mathrm{frz}}\), as an intermediate state that allows the relaxation contribution to be separated from the slab--adsorbate binding term. Here, ``intermediate state'' refers only to an algebraic state in the thermodynamic cycle. The \(X^{\mathrm{frz}}\) structure is obtained by removing the surface from a fully relaxed slab--adsorbate structure while retaining the adsorbate coordinates. It therefore represents the relaxed on-surface geometry evaluated as an isolated fragment and is not a transition-state geometry. The optimized isolated adsorbate \(X^{\mathrm{opt}}\) remains the molecular reference through the formation term. The adsorption energy is then written as
\begin{align}
\begin{split}
&\Delta E_{\mathrm{ads}}^{(2)}(X) = \\
&\Delta E_{\mathrm{bind}}^{\mathrm{QMC}}(X^{\mathrm{frz}})
+ \Delta E_{\mathrm{rlx}}^{\mathrm{HL}}(X)
+ \Delta E_{\mathrm{form}}^{\mathrm{HL}}(X^{\mathrm{opt}}),
\label{eq:scheme2_general}
\end{split}
\end{align}
where
\begin{equation}
\Delta E_{\mathrm{rlx}}(X)
= E_{X^{\mathrm{frz}}}
- E_{X^{\mathrm{opt}}}.
\label{eq:scheme2_def}
\end{equation}
This partition is retained to emphasize that Scheme~2 does not redefine the molecular reference state, but only isolates the relaxation contribution relative to the optimized isolated adsorbate.

To quantify the additional effect of geometry matching, the relaxation energy of OOH is evaluated using both SD-FNDMC and CCSD(T)/CBS:
\begin{align*}
\Delta E_{\mathrm{rlx}}^{\mathrm{QMC}}(\mathrm{OOH})
&= 0.198 \pm 0.007~\mathrm{eV},\\
\Delta E_{\mathrm{rlx}}^{\mathrm{HL}}(\mathrm{OOH})
&= 0.158~\mathrm{eV}.
\end{align*}

Among the intermediates considered here, OOH shows the largest geometry mismatch between the optimized gas-phase radical and the surface-extracted structure. In particular, the O--O bond elongates from 1.35~\AA\ in the optimized gas-phase structure to 1.43~\AA\ in the surface-extracted geometry, while the O--H bond length remains nearly unchanged and the H--O--O angle decreases from 105.0$^\circ$ to 102.1$^\circ$.

The corresponding Scheme~2 correction relative to Scheme~1 is
\begin{align}
\delta_{\mathrm{geom}}(\mathrm{OOH})
&= \Delta E_{\mathrm{ads}}^{(2)}(\mathrm{OOH})
- \Delta E_{\mathrm{ads}}^{(1)}(\mathrm{OOH})
\nonumber\\
&= \Delta E_{\mathrm{rlx}}^{\mathrm{HL}}(\mathrm{OOH})
- \Delta E_{\mathrm{rlx}}^{\mathrm{QMC}}(\mathrm{OOH})
\label{eq:delta_geom_ooh}\\
&= -0.041 \pm 0.007~\mathrm{eV}.
\nonumber
\end{align}
Thus, geometry matching produces a measurable but relatively small additional stabilization of OOH beyond the larger reference-state correction already captured by Scheme~1. Even for the most distorted intermediate considered here on Pt(111), these results suggest that reference-state imbalance contributes more strongly than geometry mismatch to the final error, while the latter enters as a secondary refinement.

Although Scheme~2 can be written for any adsorbate by extracting the frozen adsorbate geometry from the slab--adsorbate structure, it is useful only when the isolated frozen fragment remains a well-defined molecular state. This is the case for OOH, HCO, and COH, which remain identifiable radicals after the surface is removed, but not necessarily for strongly fragmented adsorbates or intermediates stabilized only by the surface. For this reason, the most robust reference cycles connect adsorbed species to chemically neighboring molecular intermediates rather than indirectly reconstructing the adsorbate from more remote reactant or product states.

\subsection{Transferability beyond Pt oxygenates: $^\ast$CHO and $^\ast$COH on Cu(111)}
These results indicate that reference-state imbalance is not specific to Pt oxygenates, but can also arise in conventional SD-FNDMC adsorption thermochemistry for chemically distinct adsorbates. This suggests that the hybrid cycle should be transferable beyond ORR and, more generally, beyond metallic oxygenate adsorption, provided that the relevant molecular benchmark remains reliable. To test this idea, we use the carbonaceous intermediates $^\ast$CHO and $^\ast$COH on Cu(111) as a chemically distinct benchmark system, reusing the previously published Cu(111) slab calculations and adsorption energetics from Ref.~\cite{fanta_CO_2025}. Here HCO denotes the isolated formyl radical corresponding to the adsorbed $^\ast$CHO intermediate. In the present work, only the additional molecular reference calculations required for the hybrid-cycle analysis were performed, namely the gas-phase HCO and COH benchmarks and the isolated frozen HCO and COH radicals extracted from the adsorbed Cu(111) geometries.

As a gas-phase anchor, the HCO formation is considered first,
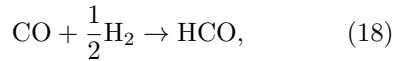
\begin{equation}
\mathrm{CO} + \frac{1}{2}\mathrm{H_2} \rightarrow \mathrm{HCO},
\end{equation}
for which accurate experimental thermochemistry is available. Table~\ref{tab:hco_coh_formation_errors} shows that the spread across density functionals remains substantial, with errors ranging from about $-0.43$ to $-0.25$~eV, whereas coupled cluster theory is essentially converged at the CCSD(T)/CBS level. In particular, CCSD(T)/CBS reproduces the experimental value within numerical uncertainty, and the remaining difference to CCSDT(Q) is below $0.01$~eV. SD-FNDMC performs significantly better than the density functionals but still underestimates the HCO formation energy by $0.043$~eV.

\begin{table}[htbp]
\caption{Gas-phase HCO formation enthalpies and COH electronic hydrogenation energies for 
$\mathrm{CO} + \frac{1}{2}\mathrm{H}_2 \rightarrow \mathrm{HCO}$ and 
$\mathrm{CO} + \frac{1}{2}\mathrm{H}_2 \rightarrow \mathrm{COH}$, in eV.
Signed errors are given in brackets, computed relative to the reference value in the bottom row of each column: the experimental 298.15~K formation enthalpy for HCO, and the CCSDT(Q) electronic hydrogenation energy for COH.}
\label{tab:hco_coh_formation_errors}
\begin{tabular*}{0.49\textwidth}{@{\extracolsep{\fill}} l c c}
\toprule\toprule
\textbf{Method} & \textbf{$dH$(HCO)} & \textbf{$dE$(COH)} \\
\midrule
PBE          & 1.152 {~[-0.426]} & 3.007 {~[-0.311]} \\
RPBE         & 1.223 {~[-0.355]} & 3.050 {~[-0.268]} \\
BEEF-vdW     & 1.308 {~[-0.270]} & 3.078 {~[-0.240]} \\
SCAN         & 1.163 {~[-0.415]} & 2.963 {~[-0.355]} \\
B3LYP        & 1.326 {~[-0.252]} & 3.102 {~[-0.216]} \\
PBE0         & 1.205 {~[-0.373]} & 2.984 {~[-0.334]} \\
CCSD         & 1.584 {~[ 0.006]} & 3.261 {~[-0.057]} \\
CCSD(T)      & 1.579 {~[ 0.001]} & 3.322 {~[ 0.004]} \\
CCSDT(Q)     & 1.572 {~[-0.006]} & 3.318 {~[ 0.000]} \\
SD-FNDMC     & 1.536 {~[-0.043]} & 3.384 {~[ 0.066]} \\
\midrule
\textbf{Exp./CCSDT(Q)} & \textbf{1.578} & \textbf{3.318} \\
\bottomrule\bottomrule
\end{tabular*}
\end{table}

COH is then treated as a complementary, theory-supported isomer. The coupled cluster convergence is very similar for HCO and COH, as shown by the electronic hydrogenation energies and signed deviations from the respective reference values reported in Table~\ref{tab:hco_coh_formation_errors}. Since HCO and COH have identical stoichiometry but distinct bonding motifs, this pair separates stoichiometric effects from electronic-structure effects in the molecular reference correction.

Using the CCSD(T)/CBS molecular reference values in Table~\ref{tab:hco_coh_formation_errors}, the corresponding Scheme~1 reference corrections are
\begin{align*}
\delta_{\mathrm{hyb}}(\mathrm{HCO}) &= +0.043 \pm 0.005~\mathrm{eV},\\
\delta_{\mathrm{hyb}}(\mathrm{COH}) &= -0.063 \pm 0.006~\mathrm{eV}.
\end{align*}
For HCO, the positive correction reflects that SD-FNDMC slightly underestimates the HCO molecular formation energy relative to the CCSD(T)/CBS benchmark. Replacing the SD-FNDMC molecular HCO formation term with a high-level reference raises the adsorption energies uniformly across sites. For COH, the negative correction reflects the opposite behavior, with SD-FNDMC overestimating the COH molecular formation energy relative to CCSD(T)/CBS, so the hybrid cycle lowers the adsorption energies. The opposite signs for HCO and COH show that the correction is not a stoichiometric offset, but is controlled by the electronic structure of the molecular reference state. 

These two corrections differ by $0.105 \pm 0.006$~eV. Because the HCO and COH formation energies are referenced to the same CO+$\frac{1}{2}$H$_2$ state, the common CO and H$_2$ contributions cancel in this difference. The differential correction therefore reflects the different SD-FNDMC errors of the HCO and COH molecular reference states despite their identical stoichiometry. This difference does not change the ordering of $^\ast$CHO and $^\ast$COH in the present Cu(111) results, but it could affect the preferred hydrogenation product when competing intermediates are close in energy. Molecular-reference corrections should therefore be evaluated separately when comparing bifurcating pathways such as $^\ast$CO hydrogenation to $^\ast$CHO or $^\ast$COH, or analogous $^\ast$NO hydrogenation pathways \cite{CalleVallejo2023}.

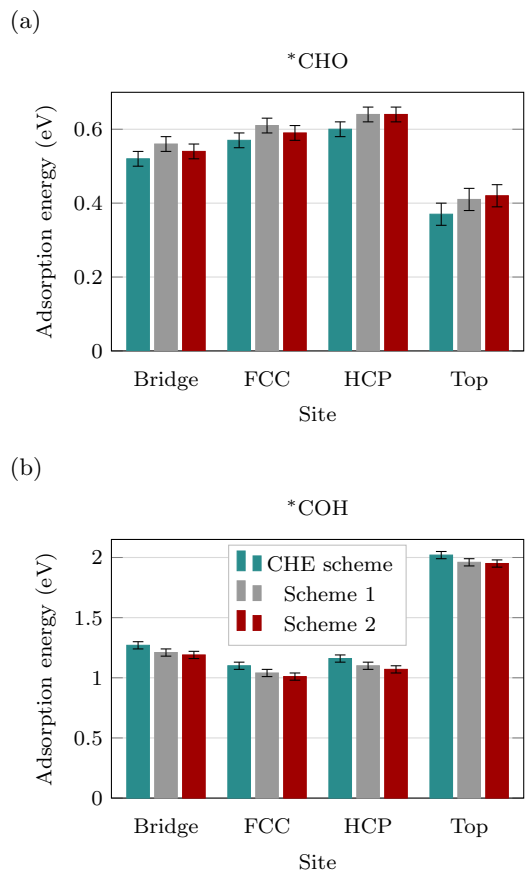
\begin{figure}[htbp]
\centering


 \definecolor{schemezero}{HTML}{298C8C}
\definecolor{schemetwo}{HTML}{A00000}

\definecolor{schemeone}{HTML}{999999}



\begin{subfigure}{\columnwidth}
    \centering
    \caption{} 
    \label{fig:hco}
    \begin{tikzpicture}
    \begin{axis}[
        width=\columnwidth,
        height=5cm,
        ybar,
        bar width=8.5pt,
        enlarge x limits=0.18,
        symbolic x coords={Bridge,FCC,HCP,Top},
        xtick=data,
        xtick style={draw=none},
        xlabel={Site},
        ymin=0,
        ymax=0.7,
        ylabel={Adsorption energy (eV)},
        title={$^\ast$CHO},
        ymajorgrids=true,
        grid style={gray!30},
        tick label style={font=\footnotesize},
        label style={font=\footnotesize},
        title style={font=\footnotesize}
    ]
    
    \addplot+[draw=schemezero, fill=schemezero, error bars/y dir=both, error bars/y explicit, error bars/error bar style={black}]
    coordinates {(Bridge,0.52) +- (0,0.02) (FCC,0.57) +- (0,0.02) (HCP,0.60) +- (0,0.02) (Top,0.37) +- (0,0.03)};
    
    \addplot+[draw=schemeone, fill=schemeone, error bars/y dir=both, error bars/y explicit, error bars/error bar style={black}]
    coordinates {(Bridge,0.56) +- (0,0.02) (FCC,0.61) +- (0,0.02) (HCP,0.64) +- (0,0.02) (Top,0.41) +- (0,0.03)};
    
    \addplot+[draw=schemetwo, fill=schemetwo, error bars/y dir=both, error bars/y explicit, error bars/error bar style={black}]
    coordinates {(Bridge,0.54) +- (0,0.02) (FCC,0.59) +- (0,0.02) (HCP,0.64) +- (0,0.02) (Top,0.42) +- (0,0.03)};
    \end{axis}
    \end{tikzpicture}
\end{subfigure}

\vspace{0.1cm} 

\begin{subfigure}{\columnwidth}
    \centering
    \caption{} 
    \label{fig:coh}
    \begin{tikzpicture}
    \begin{axis}[
        width=\columnwidth,
        height=5cm,
        ybar,
        bar width=8.5pt,
        enlarge x limits=0.18,
        symbolic x coords={Bridge,FCC,HCP,Top},
        xtick=data,
        xtick style={draw=none},
        xlabel={Site},
        ymin=0,
        ymax=2.15,
        ylabel={Adsorption energy (eV)},
        title={$^\ast$COH},
        ymajorgrids=true,
        grid style={gray!30},
        tick label style={font=\footnotesize},
        label style={font=\footnotesize},
        title style={font=\footnotesize},
        legend columns=1,
        legend style={
            draw=gray!50,
            fill=white,
            font=\footnotesize,
             at={(0.283,0.98)}, 
            anchor=north west,
        }
    ]
    
    \addplot+[draw=schemezero, fill=schemezero, error bars/y dir=both, error bars/y explicit, error bars/error bar style={black}]
    coordinates {(Bridge,1.27) +- (0,0.03) (FCC,1.10) +- (0,0.03) (HCP,1.16) +- (0,0.03) (Top,2.02) +- (0,0.03)};
    \addlegendentry{CHE scheme}
    
    \addplot+[draw=schemeone, fill=schemeone, error bars/y dir=both, error bars/y explicit, error bars/error bar style={black}]
    coordinates {(Bridge,1.21) +- (0,0.03) (FCC,1.04) +- (0,0.03) (HCP,1.10) +- (0,0.03) (Top,1.96) +- (0,0.03)};
    \addlegendentry{Scheme 1}
    
    \addplot+[draw=schemetwo, fill=schemetwo, error bars/y dir=both, error bars/y explicit, error bars/error bar style={black}]
    coordinates {(Bridge,1.19) +- (0,0.03) (FCC,1.01) +- (0,0.03) (HCP,1.07) +- (0,0.03) (Top,1.95) +- (0,0.03)};
    \addlegendentry{Scheme 2}
    \end{axis}
    \end{tikzpicture}
\end{subfigure}

\caption{
Adsorption energies of (a) $^\ast$CHO and (b) $^\ast$COH on Cu(111) at bridge, fcc, hcp, and top sites for the CHE scheme, Scheme 1, and Scheme 2. Error bars denote one standard deviation QMC statistical uncertainties.
}
\label{fig:hco_coh_barplot}
\end{figure}

The corrected adsorption energies for $^\ast$CHO and $^\ast$COH on Cu(111) are summarized in Fig.~\ref{fig:hco_coh_barplot}, with the corresponding numerical values reported in Table~S14 of the Supplemental Material. For $^\ast$CHO, the HCO molecular-reference correction in Scheme~1 increases the adsorption energy by $0.043$~eV at every site, whereas for $^\ast$COH, the COH molecular-reference correction decreases it by $0.063$~eV. These uniform site-independent shifts are inherited directly from the CCSD(T)/CBS molecular reference corrections above. The sign reversal between HCO and COH shows that the hybrid correction is not a rigid stoichiometric offset, but instead depends on the detailed electronic structure of the molecular reference state.

Scheme~2 introduces a further geometry-matched refinement by replacing the optimized gas-phase radical with the isolated radical frozen in its adsorbed geometry. As seen in Fig.~\ref{fig:hco_coh_barplot}, this additional correction is smaller than the Scheme~1 shift for both radicals. Overall, the geometry-matching correction is more site dependent for HCO than for COH. 
For COH, the Scheme~2 correction lowers the Scheme~1 adsorption energies by only $0.006$--$0.032$~eV depending on site, whereas for HCO it ranges from $-0.029$~eV at the bridge site to $+0.003$~eV at the top site. This stronger site sensitivity indicates that, unlike the nearly uniform Scheme~1 reference correction, geometry matching reflects local distortions of the adsorbed radical and is best viewed as a secondary, site-specific refinement. Overall, the Cu(111) results support the same interpretation reached for Pt oxygenates. Thus, the same reference-state correction applies beyond ORR oxygenates, but its sign and magnitude are still determined by the molecular reference state.

\subsection{Applicability, diagnostics, and limitations of the hybrid cycle}
These three constructions provide a diagnostic for when conventional SD-FNDMC adsorption thermochemistry is expected to be reliable. At the electronic-energy level, the residual error of the conventional adsorption cycle may be written schematically as
\begin{equation}
\delta_{\mathrm{ads}}^{(0)}(X) =
\underbrace{
\delta_{\mathrm{bind}}^{\mathrm{QMC}}(X)
}_{\substack{\text{slab--adsorbate} \\ \text{binding error}}}
+
\underbrace{
\delta_{\mathrm{form}}^{\mathrm{QMC}}(X)
}_{\substack{\text{molecular reference} \\ \text{error}}},
\label{eq:error_decomp_scheme0}
\end{equation}
where the first term reflects the residual error in the QMC binding contribution and the second term reflects the error in the molecular formation energy used to reference the adsorbate. This separation is formally analogous to recent DFT decompositions of adsorption energy errors into gas-phase and adsorbed-phase contributions \cite{Romeo2025,Urregoortiz2025}. However, here it is used to diagnose SD-FNDMC fixed-node error partitioning and to replace the molecular formation contribution with an explicit coupled cluster benchmark. Scheme~1 targets the second contribution by replacing $\delta_{\mathrm{form}}^{\mathrm{QMC}}(X)$ with a high-level benchmark value, while Scheme~2 tests whether adsorbate relaxation contributes an additional correction.

This decomposition also clarifies why gas-phase atomization or formation errors are useful as an initial diagnostic \cite{nemec2010,petruzielo2012,wang_performance_2019}. Large SD-FNDMC errors for isolated molecules or radicals immediately indicate that the molecular reference term in Eq.~\eqref{eq:error_decomp_scheme0} may be problematic if it enters the adsorption cycle without favorable cancellation. These quantities are indicators, not direct predictors, of adsorption energy errors. The adsorption error is controlled not by the absolute magnitude of the fixed-node bias in any one species, but by how that bias differs across the states entering the cycle. A molecule with a sizable atomization-energy error can still yield an accurate adsorption energy if the corresponding reference-state bias cancels favorably, whereas a smaller molecular error may become important if it is amplified by the thermodynamic construction, as seen above for CHE cycle formulations.

The same decomposition is not limited to adsorption on metals. It applies whenever the target thermochemical quantity can be separated into an extended-system contribution, where QMC error cancellation is expected to be favorable, and a molecular reference term that can be benchmarked independently. This includes adsorption on semiconducting or insulating surfaces, as well as reaction energies constructed from chemically distinct molecular reference states. The scope is therefore broader than metal adsorption, but it remains limited to energy differences with a well-defined molecular-reference component.

The main limitation is the loss of a well-defined detached-fragment reference. The detached fragment does not need to retain the same localized spin density, charge distribution, or formal oxidation state assignment as the adsorbate on the surface, because these quantities can change through surface coupling. It must, however, remain the same identifiable species, with unchanged stoichiometry and connectivity, so that the same total charge and spin multiplicity can be used consistently for $X^{\mathrm{frz}}$ and $X^{\mathrm{opt}}$. If the adsorbed state instead involves dissociation, rearrangement, or no corresponding well-defined isolated electronic state, the high-level molecular calculation benchmarks a different species rather than the intended reference-state contribution. The OH, OOH, HCO, and COH satisfy this criterion, and Scheme~2 is applied here to OOH, HCO, and COH.

Dissociative adsorption of O$_2$ through the reaction O$_2$(g)$+2^\ast\rightarrow2^\ast$O is not evaluated here and provides a contrasting example. Because the O--O bond is broken in the adsorbed state, intact O$_2$ cannot serve as the detached fragment for Scheme~2. As illustrated by the $^\ast$O results above, Scheme~1 can instead be formulated per adsorbed O atom using isolated atomic O as $X^{\mathrm{opt}}$, with its formation term referenced to H$_2$O and H$_2$ within the CHE construction. Scheme~2 then introduces no geometry correction because an atom has no internal structural degrees of freedom.

This reference-state limitation should be distinguished from errors in the QMC slab--adsorbate binding term. If the adsorbed state itself involves strong spin-state competition, oxidation-state changes, or near-degenerate electronic configurations, the hybrid cycle can still be written, but it will not correct residual fixed-node errors in the binding contribution. Single-atom catalysts are a prominent example: adsorbate binding can displace the metal center, change its oxidation and spin state, and strongly affect electrocatalytic energetics in M--N--C systems \cite{jung2025}. Such cases are challenging for DFT \cite{henderson2024}. The associated spin-state splittings and near-degenerate electronic states remain difficult benchmark targets and may involve stronger nondynamic correlation than the molecular references considered here \cite{reimann2022}.

The construction should not be automatically transferred to density-functional theory. Its use here relies on SD-FNDMC giving a more balanced slab--adsorbate binding term than the full adsorption cycle involving chemically dissimilar gas-phase references. In DFT, exchange-correlation error can be distributed over both the molecular reference and the adsorbate--surface interaction. The latter can be strongly functional dependent, as illustrated by the CO/Pt(111) puzzle \cite{Feibelman2001,Patra2019}. Correcting only the gas-phase formation energy does not generally isolate a robust DFT adsorption-energy component. The presented hybrid cycle is not intended as a generic post hoc correction for arbitrary approximate methods, but as a construction specifically motivated by the distinct error partitioning observed for SD-FNDMC. It should be viewed primarily as a diagnostic and corrective framework for SD-FNDMC surface thermochemistry and related reaction or adsorption energies, rather than as a general-purpose empirical correction.

\section{Conclusions}
Conventional SD-FNDMC surface thermochemistry can acquire an additional, separately identifiable bias when extended-system states are referenced to chemically dissimilar gas-phase species. This is already evident in the gas-phase oxygen thermochemistry relevant to ORR, where the CHE-derived OOH quantity exhibits a much larger residual error than direct OOH formation. Based on this diagnosis, a reference-state-balanced hybrid cycle was introduced. It retains SD-FNDMC for slab--adsorbate binding while replacing the molecular formation term by a high-level coupled cluster benchmark, thereby reducing the additional bias introduced by dissimilar molecular references. For Pt(111), this yields site-independent Scheme~1 corrections of $-0.091$~eV for O, $-0.037$~eV for OH, and $-0.183$~eV for OOH, while the geometry-matched refinement for OOH gives a further stabilization of $-0.041 \pm 0.007$~eV. The same framework transfers to $^\ast$CHO and $^\ast$COH adsorption on Cu(111), where Scheme~1 shifts the adsorption energies by $+0.043$~eV through the HCO molecular reference and by approximately $-0.063$~eV through the COH molecular reference, respectively. The corresponding Scheme~2 geometry-matching corrections are smaller and site-dependent, ranging from about $-0.029$ to $+0.003$~eV for $^\ast$CHO and from about $-0.032$ to $-0.006$~eV for $^\ast$COH. The opposite signs of the HCO and COH Scheme~1 corrections show that the hybrid correction is chemically specific rather than a rigid stoichiometric shift. The hybrid cycle identifies and reduces molecular reference bias in SD-FNDMC surface thermochemistry and in related energy differences that combine extended-system contributions with chemically distinct molecular references.

\section*{ACKNOWLEDGMENTS}
This work was performed under the Liquid Sunlight Alliance, which was supported by the U.S. Department of Energy, Office of Science, Office of Basic Energy Sciences, Fuels from Sunlight Hub under Award Number DE-SC0021266.
An award of computer time was provided by the ASCR Leadership Computing Challenge (ALCC) program for the project on High-precision heterogeneous catalysis by quantum Monte Carlo (CatalystQMC). This research used resources of the Argonne Leadership Computing Facility, which is a U.S. Department of Energy Office of Science User Facility operated under contract DE-AC02-06CH11357.
This research also used resources of the National Energy Research Scientific Computing Center, a DOE Office of Science User Facility supported by the Office of Science of the U.S. Department of Energy under Contract No. DE-AC02-05CH11231 using NERSC award BES-ERCAP0036967.

\section*{Data Availability}
The structures, molecular reference data, QMC total energies, and adsorption energies that support the findings of this article are openly available at Catalysis-Hub.org \cite{Winther2019_catalysis-hub} under the dataset entry \href{https://www.catalysis-hub.org/publications/FantaQMCRefCorrection2026}{FantaQMCRefCorrection2026}. Previously published Cu(111) structures and adsorption data reused in this work are available from Ref.~\cite{fanta_CO_2025}.

\bibliography{refs}
\clearpage
\onecolumngrid

\let\addcontentsline\oldaddcontentsline

\appendix
\setcounter{section}{0}
\setcounter{subsection}{0}
\setcounter{subsubsection}{0}
\setcounter{figure}{0}
\setcounter{table}{0}
\setcounter{equation}{0}

\renewcommand{\thefigure}{S\arabic{figure}}
\renewcommand{\thetable}{S\arabic{table}}
\renewcommand{\theequation}{S\arabic{equation}}

\section*{Supplemental Material}

\tableofcontents
\clearpage
This Supplemental Material contains the total energies and molecular reference data used to construct the gas-phase thermochemistry and hybrid adsorption energy schemes, the explicit algebraic forms of the schemes omitted from the main text, the electronic SD-FNDMC Pt(111) energies and finite-size extrapolations, and the gas-phase and Cu(111) benchmark data used for the HCO/COH transferability analysis.

\section{Molecular reference thermochemistry and total energies}
\subsection{Zero-point energies and integrated heat capacities}
To construct the 298.15~K enthalpy-like quantities used in the gas-phase benchmarks and hybrid thermodynamic cycles discussed in the main text, zero-point energies (ZPE) and integrated heat capacities were added to the electronic energies according to
\begin{equation}
\widetilde{E}_X(298.15~\mathrm{K})
=
E_X + E_X^{\mathrm{ZPE}}
+ \Delta H_X^{0\rightarrow 298.15~\mathrm{K}},
\end{equation}
Here, \(\Delta H_X^{0\rightarrow 298.15~\mathrm{K}}
=H_X^\circ(298.15~\mathrm{K})-H_X^\circ(0~\mathrm{K})\) is the integrated heat capacity from 0 to 298.15~K reported in the CCCBDB (Computational Chemistry Comparison and Benchmark DataBase). The values used for the oxygen-containing reference species relevant to ORR are summarized in Table~\ref{tab:si_corr_orr}, while the corresponding values for the carbon-containing reference species used in the Cu(111) transferability analysis are summarized in Table~\ref{tab:si_corr_carbon}. Although the individual enthalpy increments are similar, they do not cancel identically in the stoichiometric reaction combinations. Their net contributions are $-0.0302$~eV for direct H$_2$O formation, $+0.0024$~eV for direct OH formation, $-0.0303$~eV for direct OOH formation, $+0.0326$~eV for CHE-derived OH, $+0.0301$~eV for CHE-derived OOH, and $-0.0303$~eV for HCO formation. We therefore retain these enthalpy increments to construct quantities that are directly comparable with the experimental 298.15~K formation enthalpies.

\begin{table}[h]
\caption{Zero-point energies, integrated heat capacities from 0 to 298.15~K, and total 298.15~K corrections used for the oxygen-containing molecular references. All values are in eV.}
\label{tab:si_corr_orr}
\begin{tabular}{lccc}
\toprule\toprule
\textbf{Species} & \textbf{ZPE} &
\textbf{\(\Delta H^{0\rightarrow298}\)} &
\textbf{Total} \\
\midrule
H$_2$O & 0.5584 & 0.1026 & 0.6610 \\
H$_2$  & 0.2702 & 0.0878 & 0.3580 \\
O$_2$  & 0.0965 & 0.0900 & 0.1865 \\
OH     & 0.2295 & 0.0913 & 0.3208 \\
OOH    & 0.3673 & 0.1036 & 0.4709 \\
\bottomrule\bottomrule
\end{tabular}
\end{table}

No 298.15~K thermochemical correction is listed in Table~\ref{tab:si_corr_carbon} for COH because the present Cu(111) transferability analysis uses COH only in the electronic-energy comparisons (Table~\ref{tab:hco_coh_electronic_errors}), and no final benchmark vibrational/thermal correction was adopted here.

\begin{table}[h]
\caption{Zero-point energies, integrated heat capacities from 0 to 298.15~K, and total 298.15~K corrections used for the carbon-containing molecular references. All values are in eV.}
\label{tab:si_corr_carbon}
\begin{tabular}{lccc}
\toprule\toprule
\textbf{Species} & \textbf{ZPE} &
\textbf{\(\Delta H^{0\rightarrow298}\)} &
\textbf{Total} \\
\midrule
H$_2$ & 0.2702 & 0.0878 & 0.3580 \\
CO    & 0.1329 & 0.0899 & 0.2227 \\
HCO   & 0.3337 & 0.1035 & 0.4373 \\
COH   & ---    & ---    & ---    \\
\bottomrule\bottomrule
\end{tabular}
\end{table}

\clearpage
\subsection{ORR molecular reference energies}
The molecular reference energies entering the gas-phase thermochemistry and hybrid adsorption-energy schemes are summarized in Tables~\ref{tab:orr_gas_phase_total_energies} and \ref{tab:orr_reference_energies_hybrid}. Table~\ref{tab:orr_gas_phase_total_energies} reports the gas-phase total energies used to construct the benchmark formation thermochemistry across all methods, while Table~\ref{tab:orr_reference_energies_hybrid} lists the SD-FNDMC and CCSD(T)/CBS reference energies used directly in Schemes~1 and 2. The entry OOH$^{\mathrm{frz}}$ corresponds to the isolated OOH radical frozen in the geometry extracted from the adsorbed Pt(111) structure. The absolute SD-FNDMC and coupled cluster (CC) total energies should not be compared directly because the SD-FNDMC calculations use ccECP, whereas the CC calculations use all-electron Hamiltonians with frozen-core correlation. The hybrid cycles therefore use only energy differences constructed within each method.

\begin{table*}[h]
\caption{coupled cluster total energies for the gas-phase oxygen-containing reference species and $\mathrm{H}_2$, evaluated with the cc-pVDZ and cc-pVTZ basis sets. All energies are in Hartree (Eh).}
\label{tab:oxygen_cc_basis_energies}
\begin{tabular*}{\textwidth}{@{\extracolsep{\fill}} l l 
    S[table-format=-3.6] 
    S[table-format=-3.6] 
    S[table-format=-3.6] @{}}
\toprule\toprule
\textbf{System} & \textbf{Basis} & 
\multicolumn{1}{c}{\textbf{CCSD}} & 
\multicolumn{1}{c}{\textbf{CCSD(T)}} & 
\multicolumn{1}{c}{\textbf{CCSDT(Q)}} \\
\midrule
H$_2$                & cc-pVDZ &   -1.163495 & {---}       & {---}       \\
                     & cc-pVTZ &   -1.172372 & {---}       & {---}       \\
\midrule
O                    & cc-pVDZ &  -74.909271 &  -74.910020 &  -74.910199 \\
                     & cc-pVTZ &  -74.971124 &  -74.974035 &  -74.974301 \\
\midrule
H$_2$O               & cc-pVDZ &  -76.238158 &  -76.241215 &  -76.241865 \\
                     & cc-pVTZ &  -76.324641 &  -76.332322 &  -76.332756 \\
\midrule
O$_2$                & cc-pVDZ & -149.975925 & -149.985604 & -149.987705 \\
                     & cc-pVTZ & -150.111384 & -150.129151 & -150.130943 \\
\midrule
OH                   & cc-pVDZ &  -75.557636 &  -75.559404 &  -75.559857 \\
                     & cc-pVTZ &  -75.632464 &  -75.637622 &  -75.638103 \\
\midrule
OOH                  & cc-pVDZ & -150.549715 & -150.558578 & -150.560697 \\
                     & cc-pVTZ & -150.695106 & -150.712508 & -150.714410 \\
\midrule
OOH$^{\mathrm{frz}}$ & cc-pVDZ & -150.544283 & -150.554042 & -150.556875 \\
                     & cc-pVTZ & -150.688549 & -150.707090 & -150.709780 \\
\bottomrule\bottomrule
\end{tabular*}
\end{table*}

\begin{table*}[h]
\caption{Gas-phase total electronic energies for the ORR molecular references used in the thermochemical benchmark analysis. Energies are reported in the native units of each method. All coupled cluster energies are extrapolated to CBS.}
\label{tab:orr_gas_phase_total_energies}
\begin{tabular*}{\textwidth}{@{\extracolsep{\fill}} l c 
    S[table-format=-2.6] 
    S[table-format=-1.6] 
    S[table-format=-3.6] 
    S[table-format=-2.6] 
    S[table-format=-3.6] @{}}
\toprule\toprule
\textbf{Method} & \textbf{Unit} & 
\multicolumn{1}{c}{\textbf{H$_2$O}} & 
\multicolumn{1}{c}{\textbf{H$_2$}} & 
\multicolumn{1}{c}{\textbf{O$_2$}} & 
\multicolumn{1}{c}{\textbf{OH}} & 
\multicolumn{1}{c}{\textbf{OOH}} \\
\midrule
PBE            & Eh & -76.383905 & -1.166312 & -150.250318 & -75.685775 & -150.832511 \\
RPBE           & Eh & -76.464544 & -1.178714 & -150.397187 & -75.766950 & -150.982512 \\
BEEF-vdW       & eV & -12.829902 & -7.171037 &   -6.659607 &  -6.412941 &  -10.140712 \\
SCAN           & Eh & -76.437010 & -1.171442 & -150.344149 & -75.743735 & -150.928464 \\
PBE0           & Eh & -76.383638 & -1.168673 & -150.236590 & -75.688665 & -150.819191 \\
B3LYP          & Eh & -76.432861 & -1.173511 & -150.331384 & -75.738367 & -150.916019 \\
CCSD           & Eh & -76.361055 & -1.176110 & -150.168420 & -75.663971 & -150.756323 \\
CCSD(T)        & Eh & -76.370683 & -1.176110 & -150.189591 & -75.670555 & -150.777321 \\
CCSDT(Q)       & Eh & -76.371026 & -1.176110 & -150.191253 & -75.671048 & -150.778103 \\
\bottomrule\bottomrule
\end{tabular*}
\end{table*}

\begin{table*}[h]
\caption{Reference total energies used in the hybrid-cycle analysis for the ORR intermediates. SD-FNDMC values are given with one-standard-deviation statistical uncertainties. CCSD(T)/CBS values are reported for the corresponding geometries. OOH$^{\mathrm{frz}}$ denotes the isolated OOH radical frozen in the geometry extracted from Pt(111). All energies are in Hartree (Eh).}
\label{tab:orr_reference_energies_hybrid}
\begin{tabular*}{0.6\textwidth}{@{\extracolsep{\fill}} l 
    S[table-format=-2.6(6), separate-uncertainty=true] 
    S[table-format=-3.6] @{}}
\toprule\toprule
\textbf{System} & 
\multicolumn{1}{c}{\textbf{SD-FNDMC}} & 
\multicolumn{1}{c}{\textbf{CCSD(T)}} \\
\midrule
H$_2$                &  -1.175206 +- 0.000047 &   -1.176110 \\
H$_2$O               & -17.242057 +- 0.000236 &  -76.370683 \\
O$_2$                & -31.926973 +- 0.000182 & -150.189591 \\
O                    & -15.869936 +- 0.000257 &  -75.000988 \\
OH                   & -16.541035 +- 0.000240 &  -75.670555 \\
OOH$^{\mathrm{opt}}$ & -32.514694 +- 0.000207 & -150.777321 \\
OOH$^{\mathrm{frz}}$ & -32.507407 +- 0.000181 & -150.771532 \\
\bottomrule\bottomrule
\end{tabular*}
\end{table*}

\clearpage
\section{Surface ORR data and explicit scheme expressions}
\subsection{Electronic Pt(111) SD-FNDMC energies and finite-size extrapolation}
Tables~\ref{tab:elec_fndmc_combined_precise}--\ref{tab:scheme0_extrapolated_precise} summarize the electronic SD-FNDMC local energies, the corresponding cell specific CHE scheme adsorption and binding energies, and the resulting two-point \(N^{-5/4}\) extrapolated values used in the Pt(111) analysis.

\begin{table*}[h]
\caption{Total electronic SD-FNDMC local energies and variances for the base \(1\times1\) QMC cell and its \(2\times2\) tiling. Local energies are in Eh and variances in Eh\(^2\).}
\label{tab:elec_fndmc_combined_precise}
\begin{tabular*}{\textwidth}{@{\extracolsep{\fill}} l l 
    S[table-format=-4.6(6), separate-uncertainty=true] 
    S[table-format=2.6(6), separate-uncertainty=true] @{}}
\toprule\toprule
\textbf{Adsorbate block} & \textbf{Configuration} & 
{\textbf{Local Energy}} & 
{\textbf{Variance}} \\
\midrule
\multicolumn{4}{c}{\textbf{Base \(1\times1\) QMC cell}} \\
\midrule
Bare slab & empty      & -1436.069827 +- 0.000365 & 17.353527 +- 0.002842 \\
\midrule
\multirow{3}{*}{O}
& fcc        & -1452.056842 +- 0.000592 & 19.468084 +- 0.001963 \\
& hcp        & -1452.023510 +- 0.000466 & 20.689107 +- 0.476972 \\
& top        & -1451.997673 +- 0.000515 & 18.379472 +- 0.002655 \\
\midrule
\multirow{2}{*}{OH}
& fcc        & -1452.667285 +- 0.000431 & 19.178050 +- 0.002250 \\
& top        & -1452.684144 +- 0.000814 & 18.231166 +- 0.004171 \\
\midrule
OOH & top        & -1468.614804 +- 0.001129 & 18.672401 +- 0.004513 \\
\midrule
\multicolumn{4}{c}{\textbf{\(2\times2\) tiled QMC cell}} \\
\midrule
Bare slab & empty      & -5743.817772 +- 0.001017 & 55.619512 +- 0.011243 \\
\midrule
\multirow{3}{*}{O}
& fcc        & -5807.876431 +- 0.001345 & 57.217711 +- 0.009441 \\
& hcp        & -5807.740335 +- 0.000858 & 56.923147 +- 0.008626 \\
& top        & -5807.534620 +- 0.000598 & 57.302879 +- 0.010239 \\
\midrule
\multirow{2}{*}{OH}
& fcc        & -5810.275231 +- 0.000804 & 57.182875 +- 0.015678 \\
& top        & -5810.281089 +- 0.001176 & 57.004419 +- 0.015746 \\
\midrule
OOH & top        & -5874.026888 +- 0.000846 & 58.636135 +- 0.017504 \\
\bottomrule\bottomrule
\end{tabular*}
\end{table*}

\begin{table*}[h]
\caption{CHE-scheme adsorption and binding energies from SD-FNDMC for the base \(1\times1\) QMC cell and its \(2\times2\) tiling. All values are in eV.}
\label{tab:scheme0_1x1_2x2_precise}
\begin{tabular*}{\textwidth}{@{\extracolsep{\fill}} l 
    S[table-format=1.3(3), separate-uncertainty=true] 
    S[table-format=-1.3(3), separate-uncertainty=true] 
    S[table-format=1.3(3), separate-uncertainty=true] 
    S[table-format=-1.3(3), separate-uncertainty=true] @{}}
\toprule\toprule
\textbf{Adsorbate/site} & 
\multicolumn{1}{c}{\(\Delta E_{\mathrm{ads}}^{(0)}\) \((1\times1)\)} & 
\multicolumn{1}{c}{\(\Delta E_{\mathrm{bind}}\) \((1\times1)\)} & 
\multicolumn{1}{c}{\(\Delta E_{\mathrm{ads}}^{(0)}\) \((2\times2)\)} & 
\multicolumn{1}{c}{\(\Delta E_{\mathrm{bind}}\) \((2\times2)\)} \\
\midrule
O$_{\mathrm{fcc}}$   & 2.172 +- 0.020 & -3.186 +- 0.020 & 1.420 +- 0.046 & -3.938 +- 0.046 \\
O$_{\mathrm{hcp}}$   & 3.079 +- 0.017 & -2.279 +- 0.018 & 2.346 +- 0.037 & -3.012 +- 0.037 \\
O$_{\mathrm{top}}$   & 3.783 +- 0.018 & -1.576 +- 0.019 & 3.745 +- 0.033 & -1.613 +- 0.033 \\
OH$_{\mathrm{fcc}}$  & 1.551 +- 0.017 & -1.535 +- 0.017 & 1.091 +- 0.036 & -1.995 +- 0.036 \\
OH$_{\mathrm{top}}$  & 1.092 +- 0.025 & -1.994 +- 0.025 & 1.051 +- 0.043 & -2.035 +- 0.043 \\
OOH$_{\mathrm{top}}$ & 4.798 +- 0.033 & -0.824 +- 0.033 & 4.599 +- 0.037 & -1.023 +- 0.036 \\
\bottomrule\bottomrule
\end{tabular*}
\end{table*}

\begin{table}[h]
\caption{Two-point \(N^{-5/4}\) extrapolated adsorption and binding energies obtained from the \(1\times1\) and \(2\times2\) SD-FNDMC data. All values are in eV.}
\label{tab:scheme0_extrapolated_precise}
\begin{tabular}{ l 
    S[table-format=1.3(3), separate-uncertainty=true] 
    S[table-format=-1.3(3), separate-uncertainty=true] }
\toprule\toprule
\textbf{Adsorbate/site} & 
\multicolumn{1}{c}{\(\Delta E_{\mathrm{ads}}^{(0)}(\infty)\)} & 
\multicolumn{1}{c}{\(\Delta E_{\mathrm{bind}}(\infty)\)} \\
\midrule
O$_{\mathrm{fcc}}$   & 1.258 +- 0.026 & -4.100 +- 0.026 \\
O$_{\mathrm{hcp}}$   & 2.188 +- 0.023 & -3.170 +- 0.023 \\
O$_{\mathrm{top}}$   & 3.737 +- 0.023 & -1.621 +- 0.024 \\
OH$_{\mathrm{fcc}}$  & 0.992 +- 0.022 & -2.094 +- 0.022 \\
OH$_{\mathrm{top}}$  & 1.042 +- 0.032 & -2.044 +- 0.029 \\
OOH$_{\mathrm{top}}$ & 4.557 +- 0.041 & -1.065 +- 0.041 \\
\bottomrule\bottomrule
\end{tabular}
\end{table}

Uncertainties reported for adsorption, binding, and hybrid-cycle energies were propagated from the stochastic SD-FNDMC contributions only, so no statistical uncertainty was assigned to the coupled cluster reference energies.

\clearpage
\subsection{CHE scheme: explicit adsorption energy expressions}
For the oxygenated ORR intermediates considered on Pt(111), the CHE-based reference chemical potentials entering the CHE scheme in the main text are
\begin{align}
\mu_{\mathrm{OH}}^{\mathrm{CHE,QMC}}
&= E_{\mathrm{H_2O}}^{\mathrm{QMC}}
- \frac{1}{2}E_{\mathrm{H_2}}^{\mathrm{QMC}},
\label{eq:mu_oh_che_qmc_si}
\\
\mu_{\mathrm{O}}^{\mathrm{CHE,QMC}}
&= E_{\mathrm{H_2O}}^{\mathrm{QMC}}
- E_{\mathrm{H_2}}^{\mathrm{QMC}},
\label{eq:mu_o_che_qmc_si}
\\
\mu_{\mathrm{OOH}}^{\mathrm{CHE,QMC}}
&= 2E_{\mathrm{H_2O}}^{\mathrm{QMC}}
- \frac{3}{2}E_{\mathrm{H_2}}^{\mathrm{QMC}}.
\label{eq:mu_ooh_che_qmc_si}
\end{align}
The corresponding adsorption energies are
\begin{align}
\Delta E_{\mathrm{ads}}^{(0)}(\mathrm{OH})
&= E_{\ast\mathrm{OH}}^{\mathrm{QMC}} - E_{\ast}^{\mathrm{QMC}}
- \left(
E_{\mathrm{H_2O}}^{\mathrm{QMC}} - \frac{1}{2}E_{\mathrm{H_2}}^{\mathrm{QMC}}
\right),
\label{eq:scheme0_oh_si}
\\
\Delta E_{\mathrm{ads}}^{(0)}(\mathrm{O})
&= E_{\ast\mathrm{O}}^{\mathrm{QMC}} - E_{\ast}^{\mathrm{QMC}}
- \left(
E_{\mathrm{H_2O}}^{\mathrm{QMC}} - E_{\mathrm{H_2}}^{\mathrm{QMC}}
\right),
\label{eq:scheme0_o_si}
\\
\Delta E_{\mathrm{ads}}^{(0)}(\mathrm{OOH})
&=
E_{\ast\mathrm{OOH}}^{\mathrm{QMC}} - E_{\ast}^{\mathrm{QMC}}
- \left(
2E_{\mathrm{H_2O}}^{\mathrm{QMC}} - \frac{3}{2}E_{\mathrm{H_2}}^{\mathrm{QMC}}
\right).
\label{eq:scheme0_ooh_si}
\end{align}

For completeness, the corresponding direct O$_2$-based reference forms are
\begin{align}
\mu_{\mathrm{OH}}^{\mathrm{dir,QMC}}
&= \frac{1}{2}E_{\mathrm{H_2}}^{\mathrm{QMC}}
+ \frac{1}{2}E_{\mathrm{O_2}}^{\mathrm{QMC}},
\\
\mu_{\mathrm{O}}^{\mathrm{dir,QMC}} 
&= \frac{1}{2}E_{\mathrm{O_2}}^{\mathrm{QMC}},
\\
\mu_{\mathrm{OOH}}^{\mathrm{dir,QMC}}
&= \frac{1}{2}E_{\mathrm{H_2}}^{\mathrm{QMC}}
+ E_{\mathrm{O_2}}^{\mathrm{QMC}}.
\end{align}

\subsection{Scheme 1: explicit hybrid-cycle expressions}
For the ORR intermediates under the CHE reference, the high-level molecular formation terms entering Scheme~1 are
\begin{align}
\Delta E_{\mathrm{form}}^{\mathrm{HL}}(\mathrm{OH}^{\mathrm{opt}})
&= E_{\mathrm{OH}^{\mathrm{opt}}}^{\mathrm{HL}}
- \left(
E_{\mathrm{H_2O}}^{\mathrm{HL}}
- \frac{1}{2}E_{\mathrm{H_2}}^{\mathrm{HL}}
\right),
\label{eq:scheme1_form_oh_si}
\\
\Delta E_{\mathrm{form}}^{\mathrm{HL}}(\mathrm{O}^{\mathrm{opt}})
&= E_{\mathrm{O}^{\mathrm{opt}}}^{\mathrm{HL}}
- \left(
E_{\mathrm{H_2O}}^{\mathrm{HL}}
- E_{\mathrm{H_2}}^{\mathrm{HL}}
\right),
\label{eq:scheme1_form_o_si}
\\
\Delta E_{\mathrm{form}}^{\mathrm{HL}}(\mathrm{OOH}^{\mathrm{opt}})
&= E_{\mathrm{OOH}^{\mathrm{opt}}}^{\mathrm{HL}}
- \left(
2E_{\mathrm{H_2O}}^{\mathrm{HL}} - \frac{3}{2}E_{\mathrm{H_2}}^{\mathrm{HL}}
\right).
\label{eq:scheme1_form_ooh_si}
\end{align}

Substituting these expressions into the general Scheme~1 equation in the main text gives
\begin{align}
\Delta E_{\mathrm{ads}}^{(1)}(\mathrm{OH})
&= \left(
E_{\ast\mathrm{OH}}^{\mathrm{QMC}} - E_{\ast}^{\mathrm{QMC}} - E_{\mathrm{OH}^{\mathrm{opt}}}^{\mathrm{QMC}}
\right)
+ \left(
E_{\mathrm{OH}^{\mathrm{opt}}}^{\mathrm{HL}} - E_{\mathrm{H_2O}}^{\mathrm{HL}} + \frac{1}{2}E_{\mathrm{H_2}}^{\mathrm{HL}}
\right),
\label{eq:scheme1_oh_si}
\\
\Delta E_{\mathrm{ads}}^{(1)}(\mathrm{O})
&= \left(
E_{\ast\mathrm{O}}^{\mathrm{QMC}} - E_{\ast}^{\mathrm{QMC}} - E_{\mathrm{O}^{\mathrm{opt}}}^{\mathrm{QMC}}
\right)
+ \left(
E_{\mathrm{O}^{\mathrm{opt}}}^{\mathrm{HL}} - E_{\mathrm{H_2O}}^{\mathrm{HL}} + E_{\mathrm{H_2}}^{\mathrm{HL}}
\right),
\label{eq:scheme1_o_si}
\\
\Delta E_{\mathrm{ads}}^{(1)}(\mathrm{OOH})
&= \left(
E_{\ast\mathrm{OOH}}^{\mathrm{QMC}} - E_{\ast}^{\mathrm{QMC}} - E_{\mathrm{OOH}^{\mathrm{opt}}}^{\mathrm{QMC}}
\right)
+ \left(
E_{\mathrm{OOH}^{\mathrm{opt}}}^{\mathrm{HL}} - 2E_{\mathrm{H_2O}}^{\mathrm{HL}} + \frac{3}{2}E_{\mathrm{H_2}}^{\mathrm{HL}}
\right).
\label{eq:scheme1_ooh_si}
\end{align}

\subsection{Scheme~2: geometry-matched refinement for OOH}
For OOH, Scheme~2 introduces the isolated radical frozen in the geometry extracted from the adsorbed Pt(111) structure, denoted \(\mathrm{OOH}^{\mathrm{frz}}\), as an intermediate state that allows the relaxation contribution to be separated from the slab--adsorbate binding term. The optimized isolated radical \(\mathrm{OOH}^{\mathrm{opt}}\) remains the molecular reference through the formation term. The Scheme~2 adsorption energy is therefore written as
\begin{align}
\Delta E_{\mathrm{ads}}^{(2)}(\mathrm{OOH})
&= \Delta E_{\mathrm{bind}}^{\mathrm{QMC}}(\mathrm{OOH}^{\mathrm{frz}})
+ \Delta E_{\mathrm{rlx}}^{\mathrm{HL}}(\mathrm{OOH})
+ \Delta E_{\mathrm{form}}^{\mathrm{HL}}(\mathrm{OOH}^{\mathrm{opt}})
\nonumber\\
&= \left( E_{\ast\mathrm{OOH}}^{\mathrm{QMC}}
- E_{\ast}^{\mathrm{QMC}}
- E_{\mathrm{OOH}^{\mathrm{frz}}}^{\mathrm{QMC}} \right)
+ \left( E_{\mathrm{OOH}^{\mathrm{frz}}}^{\mathrm{HL}}
- E_{\mathrm{OOH}^{\mathrm{opt}}}^{\mathrm{HL}} \right)
\nonumber\\
&\qquad
+ \left( E_{\mathrm{OOH}^{\mathrm{opt}}}^{\mathrm{HL}}
- 2E_{\mathrm{H_2O}}^{\mathrm{HL}}
+ \frac{3}{2}E_{\mathrm{H_2}}^{\mathrm{HL}} \right).
\label{eq:scheme2_ooh_full}
\end{align}
Although the last two terms can be algebraically combined, we retain the partitioned form above because it makes clear that \(\mathrm{OOH}^{\mathrm{opt}}\) remains the molecular reference state, while \(\mathrm{OOH}^{\mathrm{frz}}\) is used only to isolate the relaxation contribution.

The corresponding Scheme~2 correction relative to Scheme~1 is
\begin{align}
\delta_{\mathrm{geom}}(\mathrm{OOH})
&= \Delta E_{\mathrm{ads}}^{(2)}(\mathrm{OOH})
- \Delta E_{\mathrm{ads}}^{(1)}(\mathrm{OOH})
\nonumber\\
&= \left[ E_{\ast\mathrm{OOH}}^{\mathrm{QMC}}
- E_{\ast}^{\mathrm{QMC}}
- E_{\mathrm{OOH}^{\mathrm{frz}}}^{\mathrm{QMC}}
+ E_{\mathrm{OOH}^{\mathrm{frz}}}^{\mathrm{HL}}
- 2E_{\mathrm{H_2O}}^{\mathrm{HL}}
+ \frac{3}{2}E_{\mathrm{H_2}}^{\mathrm{HL}} \right]
\nonumber\\
&\quad
- \left[ E_{\ast\mathrm{OOH}}^{\mathrm{QMC}}
- E_{\ast}^{\mathrm{QMC}}
- E_{\mathrm{OOH}^{\mathrm{opt}}}^{\mathrm{QMC}}
+ E_{\mathrm{OOH}^{\mathrm{opt}}}^{\mathrm{HL}}
- 2E_{\mathrm{H_2O}}^{\mathrm{HL}}
+ \frac{3}{2}E_{\mathrm{H_2}}^{\mathrm{HL}} \right]
\nonumber\\
&=
\left( E_{\mathrm{OOH}^{\mathrm{frz}}}^{\mathrm{HL}}
- E_{\mathrm{OOH}^{\mathrm{opt}}}^{\mathrm{HL}} \right)
- \left( E_{\mathrm{OOH}^{\mathrm{frz}}}^{\mathrm{QMC}}
- E_{\mathrm{OOH}^{\mathrm{opt}}}^{\mathrm{QMC}} \right)
\nonumber\\
&= \Delta E_{\mathrm{rlx}}^{\mathrm{HL}}(\mathrm{OOH})
- \Delta E_{\mathrm{rlx}}^{\mathrm{QMC}}(\mathrm{OOH}).
\label{eq:scheme2_ooh_correction_full}
\end{align}

Using the values reported in the main text,
\begin{align}
\Delta E_{\mathrm{rlx}}^{\mathrm{QMC}}(\mathrm{OOH})
&= 0.198 \pm 0.007~\mathrm{eV},\\
\Delta E_{\mathrm{rlx}}^{\mathrm{HL}}(\mathrm{OOH})
&= 0.158~\mathrm{eV},
\end{align}
which gives
\begin{equation}
\delta_{\mathrm{geom}}(\mathrm{OOH})
= -0.041 \pm 0.007~\mathrm{eV}.
\end{equation}

The optimized and frozen OOH reference energies entering Eqs.~\eqref{eq:scheme2_ooh_full} and \eqref{eq:scheme2_ooh_correction_full} are listed in Table~\ref{tab:orr_reference_energies_hybrid}.

\clearpage
\section{Gas-phase benchmark data for HCO and COH molecular references}
\subsection{Molecular reference energies}
Tables~\ref{tab:carbon_gas_phase_total_energies} and \ref{tab:carbon_gas_phase_fndmc} summarize the gas-phase molecular reference energies used in the HCO/COH benchmark analysis. Table~\ref{tab:carbon_gas_phase_total_energies} reports the total energies across the density-functional and coupled cluster methods considered here, while Table~\ref{tab:carbon_gas_phase_fndmc} lists the corresponding SD-FNDMC reference energies with statistical uncertainties.

\begin{table*}[h]
\caption{coupled cluster total energies for the carbon-containing gas-phase reference species and isolated HCO/COH radicals in optimized and surface extracted frozen geometries, evaluated with the cc-pVDZ and cc-pVTZ basis sets. All energies are in Hartree (Eh).}
\label{tab:carbon_cc_basis_energies}
\begin{tabular*}{\textwidth}{@{\extracolsep{\fill}} l l 
    S[table-format=-3.6] 
    S[table-format=-3.6] 
    S[table-format=-3.6] @{}}
\toprule\toprule
\textbf{System} & \textbf{Basis} & 
\multicolumn{1}{c}{\textbf{CCSD}} & 
\multicolumn{1}{c}{\textbf{CCSD(T)}} & 
\multicolumn{1}{c}{\textbf{CCSDT(Q)}} \\
\midrule
H$_2$                   & cc-pVDZ &   -1.163495 & {---}       & {---}       \\
                        & cc-pVTZ &   -1.172372 & {---}       & {---}       \\
\midrule
CO                      & cc-pVDZ & -113.043854 & -113.054567 & -113.056118 \\
                        & cc-pVTZ & -113.138625 & -113.155602 & -113.156854 \\
\midrule
COH                     & cc-pVDZ & -113.501628 & -113.510086 & -113.511581 \\
                        & cc-pVTZ & -113.603756 & -113.618487 & -113.619817 \\
\midrule
HCO                     & cc-pVDZ & -113.565574 & -113.576049 & -113.577861 \\
                        & cc-pVTZ & -113.666982 & -113.684027 & -113.685528 \\
\midrule
COH$_{\mathrm{bridge}}$ & cc-pVDZ & -113.501146 & -113.509849 & -113.511424 \\
                        & cc-pVTZ & -113.602580 & -113.617574 & -113.618980 \\
\midrule
COH$_{\mathrm{fcc}}$    & cc-pVDZ & -113.500366 & -113.509150 & -113.510755 \\
                        & cc-pVTZ & -113.601671 & -113.616758 & -113.618191 \\
\midrule
COH$_{\mathrm{hcp}}$    & cc-pVDZ & -113.500405 & -113.509189 & -113.510793 \\
                        & cc-pVTZ & -113.601705 & -113.616792 & -113.618224 \\
\midrule
COH$_{\mathrm{top}}$    & cc-pVDZ & -113.502404 & -113.510908 & -113.512414 \\
                        & cc-pVTZ & -113.604245 & -113.619008 & -113.620352 \\
\midrule
HCO$_{\mathrm{bridge}}$ & cc-pVDZ & -113.559320 & -113.571318 & -113.573548 \\
                        & cc-pVTZ & -113.659240 & -113.678066 & -113.679959 \\
\midrule
HCO$_{\mathrm{fcc}}$    & cc-pVDZ & -113.562300 & -113.573896 & -113.575976 \\
                        & cc-pVTZ & -113.662352 & -113.680700 & -113.682448 \\
\midrule
HCO$_{\mathrm{hcp}}$    & cc-pVDZ & -113.562502 & -113.574059 & -113.576123 \\
                        & cc-pVTZ & -113.662478 & -113.680780 & -113.682511 \\
\midrule
HCO$_{\mathrm{top}}$    & cc-pVDZ & -113.564077 & -113.575272 & -113.577270 \\
                        & cc-pVTZ & -113.664777 & -113.682671 & -113.684344 \\
\bottomrule\bottomrule
\end{tabular*}
\end{table*}

\begin{table*}[h]
\caption{Gas-phase total energies for the carbon-containing reference species used in the molecular benchmark analysis. Energies are reported in the native units of each method. All coupled cluster energies are extrapolated to CBS.}
\label{tab:carbon_gas_phase_total_energies}
\begin{tabular*}{\textwidth}{@{\extracolsep{\fill}} l c 
    S[table-format=-1.6] 
    S[table-format=-3.6] 
    S[table-format=-3.6] 
    S[table-format=-3.6] @{}}
\toprule\toprule
\textbf{Method} & \textbf{Unit} & 
\multicolumn{1}{c}{\textbf{H$_2$}} & 
\multicolumn{1}{c}{\textbf{CO}} & 
\multicolumn{1}{c}{\textbf{HCO}} & 
\multicolumn{1}{c}{\textbf{COH}} \\
\midrule
PBE             & Eh & -1.166312 & -113.235140 & -113.777272 & -113.707787 \\
RPBE            & Eh & -1.178714 & -113.365153 & -113.910873 & -113.842442 \\
BEEF-vdW        & eV & -7.171037 &  -12.088328 &  -14.401390 &  -12.595593 \\
SCAN            & Eh & -1.171442 & -113.314630 & -113.858929 & -113.791464 \\
B3LYP           & Eh & -1.173511 & -113.311254 & -113.850601 & -113.783995 \\
PBE0            & Eh & -1.168673 & -113.231532 & -113.772900 & -113.706209 \\
CCSD            & Eh & -1.176110 & -113.178528 & -113.709680 & -113.646758 \\
CCSD(T)         & Eh & -1.176110 & -113.198144 & -113.729492 & -113.664129 \\
CCSDT(Q)        & Eh & -1.176110 & -113.199270 & -113.730861 & -113.665390 \\
\bottomrule\bottomrule
\end{tabular*}
\end{table*}

\begin{table}[h] 
\caption{Gas-phase total energies for the carbon-containing reference species computed with SD-FNDMC. Energies are reported in Hartree (Eh) with one-standard-deviation statistical uncertainties.}
\label{tab:carbon_gas_phase_fndmc}
\begin{tabular}{ l S[table-format=-2.6(3), separate-uncertainty=true] }
\toprule\toprule
\textbf{Species} & \multicolumn{1}{c}{\textbf{Total Energy}} \\
\midrule
H$_2$ &  -1.175115(45) \\
CO    & -21.690191(95) \\
HCO   & -22.222606(145) \\
COH   & -22.153373(185) \\
\bottomrule\bottomrule
\end{tabular}
\end{table}

\clearpage
\subsection{Electronic hydrogenation energies}
To assess the quality of the molecular reference energies used in the Cu(111) transferability analysis, Table~\ref{tab:hco_coh_electronic_errors} summarizes the electronic hydrogenation energies for HCO and COH formation together with signed errors relative to the CCSDT(Q) reference evaluated at the same geometries. 

\begin{table*}[h]
\caption{Electronic hydrogenation energies $dE$ for HCO and COH formation,
$\mathrm{CO} + \frac{1}{2}\mathrm{H}_2 \rightarrow \mathrm{HCO}$ and
$\mathrm{CO} + \frac{1}{2}\mathrm{H}_2 \rightarrow \mathrm{COH}$, in eV.
Signed errors are given relative to the CCSDT(Q) reference at the same geometries,
$\varepsilon = dE^{\mathrm{calc}} - dE^{\mathrm{CCSDT(Q)}}.$
}
\label{tab:hco_coh_electronic_errors}
\begin{tabular*}{\textwidth}{@{\extracolsep{\fill}} l
S[table-format=1.3]
S[table-format=-1.3]
S[table-format=1.3]
S[table-format=-1.3] @{}}
\toprule\toprule
& \multicolumn{2}{c}{\textbf{HCO hydrogenation}} & \multicolumn{2}{c}{\textbf{COH hydrogenation}} \\
\cmidrule(lr){2-3} \cmidrule(lr){4-5}
\textbf{Method}
& \multicolumn{1}{c}{$dE$}
& \multicolumn{1}{c}{Error}
& \multicolumn{1}{c}{$dE$}
& \multicolumn{1}{c}{Error} \\
\midrule
PBE             & 1.116 & -0.420 & 3.007 & -0.311 \\
RPBE            & 1.187 & -0.349 & 3.050 & -0.268 \\
BEEF-vdW        & 1.272 & -0.264 & 3.078 & -0.240 \\
SCAN            & 1.127 & -0.409 & 2.963 & -0.355 \\
B3LYP           & 1.290 & -0.246 & 3.102 & -0.216 \\
PBE0            & 1.169 & -0.367 & 2.984 & -0.334 \\
CCSD            & 1.548 &  0.012 & 3.261 & -0.057 \\
CCSD(T)         & 1.543 &  0.007 & 3.322 &  0.004 \\
CCSDT(Q)        & 1.536 &  0.000 & 3.318 &  0.000 \\
SD-FNDMC        & 1.501(5) & -0.035 & 3.384(6) &  0.066 \\
\bottomrule\bottomrule
\end{tabular*}
\end{table*}

\subsection{Site-specific coupled cluster reference energies}
Table~\ref{tab:hco_coh_cc_total_energies} reports the coupled cluster total energies of the optimized and adsorbate-site-specific frozen molecular reference states used to assess the HCO and COH molecular benchmarks entering the Cu(111) transferability analysis. The site-specific HCO and COH entries listed below correspond to isolated radicals frozen in the geometries extracted from the respective Cu(111) adsorption structures reported in Ref.~\cite{fanta_CO_2025}.

\begin{table*}[h]
\caption{Site-specific coupled cluster CBS total energies for isolated HCO and COH radicals frozen in geometries extracted from Cu(111) adsorption structures, together with the corresponding optimized gas-phase reference species. All energies are in Hartree (Eh).}
\label{tab:hco_coh_cc_total_energies}
\begin{tabular*}{\textwidth}{@{\extracolsep{\fill}} l 
    S[table-format=-3.6] 
    S[table-format=-3.6] 
    S[table-format=-3.6] @{}}
\toprule\toprule
\textbf{System} & 
\multicolumn{1}{c}{\textbf{CCSD}} & 
\multicolumn{1}{c}{\textbf{CCSD(T)}} & 
\multicolumn{1}{c}{\textbf{CCSDT(Q)}} \\
\midrule
COH$_{\mathrm{bridge}}$ & -113.645290 & -113.662932 & -113.664266 \\
COH$_{\mathrm{fcc}}$    & -113.644326 & -113.662067 & -113.663427 \\
COH$_{\mathrm{hcp}}$    & -113.644358 & -113.662099 & -113.663459 \\
COH$_{\mathrm{top}}$    & -113.647126 & -113.664524 & -113.665800 \\
HCO$_{\mathrm{bridge}}$ & -113.701311 & -113.723012 & -113.724764 \\
HCO$_{\mathrm{fcc}}$    & -113.704479 & -113.725671 & -113.727278 \\
HCO$_{\mathrm{hcp}}$    & -113.704574 & -113.725715 & -113.727305 \\
HCO$_{\mathrm{top}}$    & -113.707177 & -113.727892 & -113.729428 \\
\midrule
COH                     & -113.646758 & -113.664129 & -113.665390 \\
HCO                     & -113.709680 & -113.729492 & -113.730861 \\
H$_2$                   &   -1.176110 &   -1.176110 &   -1.176110 \\
CO                      & -113.178528 & -113.198144 & -113.199270 \\
\bottomrule\bottomrule
\end{tabular*}
\end{table*}

\clearpage
\section{Cu(111) adsorption and binding energies for $^\ast$CHO and $^\ast$COH}
\subsection{Final adsorption energies from the CHE and hybrid schemes}
The Cu(111) transferability analysis reuses the previously published slab geometries and SD-FNDMC adsorption data for $^\ast$CO, $^\ast$COH, and $^\ast$CHO from Ref.~\cite{fanta_CO_2025}. Here, HCO denotes the isolated formyl radical corresponding to the adsorbed CHO* intermediate. In the present work, only the additional molecular reference calculations required to construct Schemes~1 and 2 were carried out, namely the gas-phase HCO and COH benchmarks and the isolated HCO and COH radicals frozen in the geometries extracted from the adsorbed Cu(111) structures. The original Cu(111) surface geometries and electronic adsorption data are available in Ref.~\cite{fanta_CO_2025}.

Table~\ref{tab:hco_coh_schemes} shows the final adsorption energies for $^\ast$CHO and $^\ast$COH on Cu(111) obtained from the three electronic-energy constructions discussed in the main text, allowing the overall effect of the reference-state and geometry-matching corrections to be compared directly across sites and intermediates.

\begin{table*}[h]
\caption{Adsorption energies for $^\ast$CHO and $^\ast$COH on Cu(111) obtained with the three adsorption energy constructions discussed in the main text. Values are given in eV with one-standard-deviation statistical uncertainties from SD-FNDMC. The CHE scheme corresponds to conventional SD-FNDMC adsorption thermochemistry, Scheme~1 to the reference-state-balanced hybrid cycle, and Scheme~2 to the geometry-matched hybrid cycle.}
\label{tab:hco_coh_schemes}
\begin{tabular*}{\textwidth}{@{\extracolsep{\fill}}
l
S[table-format=1.3(3), separate-uncertainty=true]
S[table-format=1.3(3), separate-uncertainty=true]
S[table-format=1.3(3), separate-uncertainty=true]
S[table-format=1.3(3), separate-uncertainty=true]
S[table-format=1.3(3), separate-uncertainty=true]
S[table-format=1.3(3), separate-uncertainty=true]
}
\toprule\toprule
& \multicolumn{3}{c}{\textbf{$^\ast$CHO adsorption energy}}
& \multicolumn{3}{c}{\textbf{$^\ast$COH adsorption energy}} \\
\cmidrule(lr){2-4}\cmidrule(lr){5-7}
\textbf{Site} 
& {\textbf{CHE scheme}} 
& {\textbf{Scheme 1}} 
& {\textbf{Scheme 2}}
& {\textbf{CHE scheme}} 
& {\textbf{Scheme 1}} 
& {\textbf{Scheme 2}} \\
\midrule
bridge & 0.522 +- 0.022 & 0.564 +- 0.022 & 0.535 +- 0.023 & 1.274 +- 0.030 & 1.212 +- 0.030 & 1.191 +- 0.030\\
fcc    & 0.567 +- 0.021 & 0.609 +- 0.022 & 0.593 +- 0.021 & 1.104 +- 0.026 & 1.041 +- 0.027 & 1.013 +- 0.026\\
hcp    & 0.601 +- 0.021 & 0.643 +- 0.022 & 0.636 +- 0.022 & 1.162 +- 0.028 & 1.099 +- 0.028 & 1.068 +- 0.028\\
top    & 0.368 +- 0.025 & 0.410 +- 0.026 & 0.414 +- 0.026 & 2.021 +- 0.031 & 1.958 +- 0.032 & 1.952 +- 0.031\\
\bottomrule\bottomrule
\end{tabular*}
\end{table*}

\subsection{Finite-size analysis of $^\ast$CHO adsorption and binding energies}
Tables~\ref{tab:hco_ads_scheme0_cells}--\ref{tab:elec_fndmc_hco_precise} summarize the adsorption and binding energies for the \(1\times1\) and \(2\times2\) QMC cells, the corresponding finite-size extrapolations, and the underlying electronic SD-FNDMC energies used for the $^\ast$CHO/Cu(111) analysis.

\begin{table*}[h]
\caption{CHE scheme adsorption energies for $^\ast$CHO on Cu(111), obtained using the CHE reference construction for the \(1\times1\) and \(2\times2\) cells, together with the corresponding two-point \(N^{-5/4}\) extrapolated values. All energies are in eV.}
\label{tab:hco_ads_scheme0_cells}
\begin{tabular*}{\textwidth}{@{\extracolsep{\fill}} l 
    S[table-format=-1.3(3), separate-uncertainty=true] 
    S[table-format=-1.3(3), separate-uncertainty=true] 
    S[table-format=-1.3(3), separate-uncertainty=true] @{}}
\toprule\toprule
\textbf{Site} & 
\multicolumn{1}{c}{\textbf{\(\Delta E_{\mathrm{ads}}^{(0)}\) \((2\times2)\)}} & 
\multicolumn{1}{c}{\textbf{\(\Delta E_{\mathrm{ads}}^{(0)}\) \((1\times1)\)}} & 
\multicolumn{1}{c}{\textbf{2-point extrapolation}} \\
\midrule
bridge & 0.589 +- 0.010 & 0.905 +- 0.019 & 0.522 +- 0.022 \\
fcc    & 0.640 +- 0.011 & 0.984 +- 0.018 & 0.567 +- 0.021 \\
hcp    & 0.671 +- 0.011 & 1.001 +- 0.018 & 0.600 +- 0.021 \\
top    & 0.477 +- 0.010 & 0.988 +- 0.023 & 0.368 +- 0.025 \\
\bottomrule\bottomrule
\end{tabular*}
\end{table*}

\begin{table*}[h]
\caption{Binding energies for HCO on Cu(111) using the optimized isolated HCO geometry, corresponding to the binding term entering Scheme~1. Values for the \(1\times1\) and \(2\times2\) cells are shown together with the corresponding two-point \(N^{-5/4}\) extrapolated values. All energies are in eV.}
\label{tab:hco_bind_opt_cells}
\begin{tabular*}{\textwidth}{@{\extracolsep{\fill}} l 
    S[table-format=-1.3(3), separate-uncertainty=true] 
    S[table-format=-1.3(3), separate-uncertainty=true] 
    S[table-format=-1.3(3), separate-uncertainty=true] @{}}
\toprule\toprule
\textbf{Site} & 
\multicolumn{1}{c}{\textbf{\(\Delta E_{\mathrm{bind}}(X^{\mathrm{opt}})\) \((2\times2)\)}} & 
\multicolumn{1}{c}{\textbf{\(\Delta E_{\mathrm{bind}}(X^{\mathrm{opt}})\) \((1\times1)\)}} & 
\multicolumn{1}{c}{\textbf{2-point extrapolation}} \\
\midrule
bridge & -0.911 +- 0.011 & -0.596 +- 0.020 & -0.979 +- 0.022 \\
fcc    & -0.860 +- 0.012 & -0.517 +- 0.018 & -0.934 +- 0.022 \\
hcp    & -0.830 +- 0.012 & -0.499 +- 0.018 & -0.901 +- 0.022 \\
top    & -1.023 +- 0.011 & -0.513 +- 0.023 & -1.133 +- 0.026 \\
\bottomrule\bottomrule
\end{tabular*}
\end{table*}

\begin{table*}[h]
\caption{Binding energies for HCO on Cu(111) using the isolated radical frozen in its adsorbed geometry, corresponding to the binding term entering Scheme~2. Values for the \(1\times1\) and \(2\times2\) cells are shown together with the corresponding two-point \(N^{-5/4}\) extrapolated values. All energies are in eV.}
\label{tab:hco_bind_frz_cells}
\begin{tabular*}{\textwidth}{@{\extracolsep{\fill}} l 
    S[table-format=-1.3(3), separate-uncertainty=true] 
    S[table-format=-1.3(3), separate-uncertainty=true] 
    S[table-format=-1.3(3), separate-uncertainty=true] @{}}
\toprule\toprule
\textbf{Site} & 
\multicolumn{1}{c}{\textbf{\(\Delta E_{\mathrm{bind}}(X^{\mathrm{frz}})\) \((2\times2)\)}} & 
\multicolumn{1}{c}{\textbf{\(\Delta E_{\mathrm{bind}}(X^{\mathrm{frz}})\) \((1\times1)\)}} & 
\multicolumn{1}{c}{\textbf{2-point extrapolation}} \\
\midrule
bridge & -1.116 +- 0.011 & -0.801 +- 0.020 & -1.184 +- 0.023 \\
fcc    & -0.980 +- 0.012 & -0.637 +- 0.018 & -1.054 +- 0.021 \\
hcp    & -0.939 +- 0.012 & -0.608 +- 0.018 & -1.010 +- 0.022 \\
top    & -1.063 +- 0.011 & -0.553 +- 0.023 & -1.173 +- 0.026 \\
\bottomrule\bottomrule
\end{tabular*}
\end{table*}

For completeness, Table~\ref{tab:elec_fndmc_hco_precise} reports the electronic SD-FNDMC local energies and variances for the \(1\times1\) and \(2\times2\) HCO/Cu(111) calculations together with the corresponding gas-phase reference calculations used to construct the adsorption and binding energies above.

\begin{table*}[h]
\caption{Total electronic SD-FNDMC local energies and variances for the conventional \(1\times1\) and tiled \(2\times2\) cells with HCO adsorbate, along with corresponding gas-phase molecular references. Local energies are in Eh and variances in Eh\(^2\).}
\label{tab:elec_fndmc_hco_precise}
\begin{tabular*}{\textwidth}{@{\extracolsep{\fill}} l l 
    S[table-format=-4.6(6), separate-uncertainty=true] 
    S[table-format=2.6(6), separate-uncertainty=true] @{}}
\toprule\toprule
\textbf{System / Block} & \textbf{Configuration} & 
{\textbf{Local Energy}} & 
{\textbf{Variance}} \\
\midrule
\multicolumn{4}{c}{\textbf{Base \(1\times1\) QMC cell}} \\
\midrule
Bare slab & clean slab & -1765.470782 +- 0.000569 & 27.241337 +- 0.003844 \\
\midrule
\multirow{4}{*}{HCO}
& bridge     & -1787.715289 +- 0.000421 & 29.014778 +- 0.001533 \\
& fcc        & -1787.712383 +- 0.000311 & 29.080840 +- 0.001882 \\
& hcp        & -1787.711731 +- 0.000333 & 28.884998 +- 0.002434 \\
& top        & -1787.712234 +- 0.000618 & 29.495782 +- 0.001825 \\
\midrule
\multicolumn{4}{c}{\textbf{\(2\times2\) tiled QMC cell}} \\
\midrule
Bare slab & clean slab & -7061.777628 +- 0.001003 & 95.511731 +- 0.008518 \\
\midrule
\multirow{4}{*}{HCO}
& bridge     & -7150.801978 +- 0.001030 & 98.409701 +- 0.010984 \\
& fcc        & -7150.794473 +- 0.001282 & 98.715404 +- 0.011646 \\
& hcp        & -7150.789997 +- 0.001254 & 98.391657 +- 0.020168 \\
& top        & -7150.818445 +- 0.001078 & 98.935753 +- 0.010250 \\
\midrule
\multicolumn{4}{c}{\textbf{Molecular References}} \\
\midrule
CO        & isolated   &   -21.690191 +- 0.000095 &  0.438766 +- 0.000422 \\
H$_2$     & isolated   &    -1.175115 +- 0.000045 &  0.010035 +- 0.000010 \\
\midrule
\multirow{5}{*}{HCO}
& opt        &   -22.222606 +- 0.000145 &  0.413124 +- 0.000992 \\
& bridge frz &   -22.215060 +- 0.000197 &  0.432295 +- 0.000738 \\
& fcc frz    &   -22.218183 +- 0.000116 &  0.438485 +- 0.000637 \\
& hcp frz    &   -22.218593 +- 0.000146 &  0.439074 +- 0.000675 \\
& top frz    &   -22.221125 +- 0.000156 &  0.436566 +- 0.000688 \\
\bottomrule\bottomrule
\end{tabular*}
\end{table*}

\clearpage
\subsection{Finite-size analysis of $^\ast$COH adsorption and binding energies}
Tables~\ref{tab:coh_ads_scheme0_cells}--\ref{tab:elec_fndmc_coh_precise} summarize the adsorption and binding energies for the \(1\times1\) and \(2\times2\) QMC cells, the corresponding finite-size extrapolations, and the underlying electronic SD-FNDMC energies used for the $^\ast$COH/Cu(111) analysis.

\begin{table*}[h]
\caption{CHE scheme adsorption energies for $^\ast$COH on Cu(111), obtained using the CHE reference construction for the \(1\times1\) and \(2\times2\) cells, together with the corresponding two-point \(N^{-5/4}\) extrapolated values. All energies are in eV.}
\label{tab:coh_ads_scheme0_cells}
\begin{tabular*}{\textwidth}{@{\extracolsep{\fill}} l 
    S[table-format=1.3(3), separate-uncertainty=true] 
    S[table-format=1.3(3), separate-uncertainty=true] 
    S[table-format=1.3(3), separate-uncertainty=true] @{}}
\toprule\toprule
\textbf{Site} & 
\multicolumn{1}{c}{\textbf{\(\Delta E_{\mathrm{ads}}^{(0)}\) \((2\times2)\)}} & 
\multicolumn{1}{c}{\textbf{\(\Delta E_{\mathrm{ads}}^{(0)}\) \((1\times1)\)}} & 
\multicolumn{1}{c}{\textbf{2-point extrapolation}} \\
\midrule
bridge & 1.332 +- 0.011 & 1.603 +- 0.028 & 1.274 +- 0.030 \\
fcc    & 1.138 +- 0.013 & 1.300 +- 0.023 & 1.104 +- 0.026 \\
hcp    & 1.194 +- 0.011 & 1.342 +- 0.025 & 1.162 +- 0.028 \\
top    & 2.250 +- 0.011 & 3.316 +- 0.029 & 2.021 +- 0.031 \\
\bottomrule\bottomrule
\end{tabular*}
\end{table*}

\begin{table*}[h]
\caption{Binding energies for COH on Cu(111) using the optimized isolated COH geometry, corresponding to the binding term entering Scheme~1. Values for the \(1\times1\) and \(2\times2\) cells are shown together with the corresponding two-point \(N^{-5/4}\) extrapolated values. All energies are in eV.}
\label{tab:coh_bind_opt_cells}
\begin{tabular*}{\textwidth}{@{\extracolsep{\fill}} l 
    S[table-format=-1.3(3), separate-uncertainty=true] 
    S[table-format=-1.3(3), separate-uncertainty=true] 
    S[table-format=-1.3(3), separate-uncertainty=true] @{}}
\toprule\toprule
\textbf{Site} & 
\multicolumn{1}{c}{\textbf{\(\Delta E_{\mathrm{bind}}(X^{\mathrm{opt}})\) \((2\times2)\)}} & 
\multicolumn{1}{c}{\textbf{\(\Delta E_{\mathrm{bind}}(X^{\mathrm{opt}})\) \((1\times1)\)}} & 
\multicolumn{1}{c}{\textbf{2-point extrapolation}} \\
\midrule
bridge & -2.052 +- 0.011 & -1.782 +- 0.028 & -2.110 +- 0.015 \\
fcc    & -2.246 +- 0.013 & -2.084 +- 0.023 & -2.281 +- 0.017 \\
hcp    & -2.191 +- 0.012 & -2.043 +- 0.025 & -2.223 +- 0.016 \\
top    & -1.135 +- 0.012 & -0.069 +- 0.029 & -1.364 +- 0.015 \\
\bottomrule\bottomrule
\end{tabular*}
\end{table*}

\begin{table*}[h]
\caption{Binding energies for COH on Cu(111) using the isolated radical frozen in its adsorbed geometry, corresponding to the binding term entering Scheme~2. Values for the \(1\times1\) and \(2\times2\) cells are shown together with the corresponding two-point \(N^{-5/4}\) extrapolated values. All energies are in eV.}
\label{tab:coh_bind_frz_cells}
\begin{tabular*}{\textwidth}{@{\extracolsep{\fill}} l 
    S[table-format=-1.3(3), separate-uncertainty=true] 
    S[table-format=-1.3(3), separate-uncertainty=true] 
    S[table-format=-1.3(3), separate-uncertainty=true] @{}}
\toprule\toprule
\textbf{Site} & 
\multicolumn{1}{c}{\textbf{\(\Delta E_{\mathrm{bind}}(X^{\mathrm{frz}})\) \((2\times2)\)}} & 
\multicolumn{1}{c}{\textbf{\(\Delta E_{\mathrm{bind}}(X^{\mathrm{frz}})\) \((1\times1)\)}} & 
\multicolumn{1}{c}{\textbf{2-point extrapolation}} \\
\midrule
bridge & -2.106 +- 0.011 & -1.835 +- 0.028 & -2.164 +- 0.015 \\
fcc    & -2.330 +- 0.013 & -2.168 +- 0.023 & -2.364 +- 0.017 \\
hcp    & -2.278 +- 0.012 & -2.130 +- 0.025 & -2.309 +- 0.015 \\
top    & -1.130 +- 0.011 & -0.063 +- 0.029 & -1.359 +- 0.015 \\
\bottomrule\bottomrule
\end{tabular*}
\end{table*}

For completeness, Table~\ref{tab:elec_fndmc_coh_precise} reports the electronic SD-FNDMC local energies and variances for the \(1\times1\) and \(2\times2\) COH/Cu(111) calculations together with the corresponding gas-phase reference calculations used to construct the adsorption and binding energies above.

\begin{table*}[h]
\caption{Total electronic SD-FNDMC local energies and variances for the conventional \(1\times1\) and tiled \(2\times2\) cells with COH adsorbate, along with corresponding gas-phase molecular references. Local energies are in Eh and variances in Eh\(^2\).}
\label{tab:elec_fndmc_coh_precise}
\begin{tabular*}{\textwidth}{@{\extracolsep{\fill}} l l 
    S[table-format=-4.6(6), separate-uncertainty=true] 
    S[table-format=2.6(6), separate-uncertainty=true] @{}}
\toprule\toprule
\textbf{System / Block} & \textbf{Configuration} & 
{\textbf{Local Energy}} & 
{\textbf{Variance}} \\
\midrule
\multicolumn{4}{c}{\textbf{Base \(1\times1\) QMC cell}} \\
\midrule
Bare slab & clean slab & -1765.470782 +- 0.000569 & 27.241337 +- 0.003844 \\
\midrule
\multirow{4}{*}{COH}
& bridge     & -1787.689634 +- 0.000840 & 28.682599 +- 0.003471 \\
& fcc        & -1787.700742 +- 0.000603 & 28.195801 +- 0.003570 \\
& hcp        & -1787.699223 +- 0.000718 & 28.569572 +- 0.004876 \\
& top        & -1787.626676 +- 0.000900 & 29.175561 +- 0.003694 \\
\midrule
\multicolumn{4}{c}{\textbf{\(2\times2\) tiled QMC cell}} \\
\midrule
Bare slab & clean slab & -7061.777628 +- 0.001003 & 95.511731 +- 0.008518 \\
\midrule
\multirow{4}{*}{COH}
& bridge     & -7150.692755 +- 0.001139 & 97.597996 +- 0.013712 \\
& fcc        & -7150.721289 +- 0.001487 & 98.746642 +- 0.009125 \\
& hcp        & -7150.713159 +- 0.001284 & 98.328217 +- 0.012003 \\
& top        & -7150.557933 +- 0.001170 & 97.782089 +- 0.010262 \\
\midrule
\multicolumn{4}{c}{\textbf{Molecular References}} \\
\midrule
CO        & isolated   &   -21.690191 +- 0.000095 &  0.438766 +- 0.000422 \\
H$_2$     & isolated   &    -1.175115 +- 0.000045 &  0.010035 +- 0.000010 \\
\midrule
\multirow{5}{*}{COH}
& opt        &   -22.153373 +- 0.000185 &  0.399187 +- 0.000501 \\
& bridge frz &   -22.151402 +- 0.000168 &  0.429304 +- 0.000581 \\
& fcc frz    &   -22.150301 +- 0.000170 &  0.429965 +- 0.000491 \\
& hcp frz    &   -22.150183 +- 0.000141 &  0.427314 +- 0.000457 \\
& top frz    &   -22.153566 +- 0.000137 &  0.431182 +- 0.000548 \\
\bottomrule\bottomrule
\end{tabular*}
\end{table*}

\end{document}